\documentclass[reprint,superscriptaddress, amsmath,amssymb, prl,]{revtex4-2}

\usepackage{graphicx}
\usepackage{subfigure}
\usepackage{dcolumn}
\usepackage{bm}
\usepackage[colorlinks, linkcolor=red, anchorcolor=blue, citecolor=blue,urlcolor=black]{hyperref}
\usepackage{ulem}
\usepackage{verbatim}
\usepackage{mathrsfs}

\newcommand{\beq}{\begin{eqnarray} }
\newcommand{\eeq}{\end{eqnarray} }
\newcommand{\Beq}{\begin{eqnarray*} }
\newcommand{\Eeq}{\end{eqnarray*} }
\newcommand{\Bmat}{\left(\begin{matrix}}
\newcommand{\Emat}{\end{matrix}\right)}
\newcommand{\up}{\uparrow}
\newcommand{\dn}{\downarrow}

\begin{document}


\title{Quantum Anomalous Hall Effect in Antiferromagnetism}

\author{Peng-Jie Guo}\email{guopengjie@ruc.edu.cn}
\affiliation{Department of Physics and Beijing Key Laboratory of Opto-electronic Functional Materials $\&$ Micro-nano Devices, Renmin University of China, Beijing 100872, China}
\author{Zheng-Xin Liu}\email{liuzxphys@ruc.edu.cn}
\affiliation{Department of Physics and Beijing Key Laboratory of Opto-electronic Functional Materials $\&$ Micro-nano Devices, Renmin University of China, Beijing 100872, China}
\author{Zhong-Yi Lu}\email{zlu@ruc.edu.cn}
\affiliation{Department of Physics and Beijing Key Laboratory of Opto-electronic Functional Materials $\&$ Micro-nano Devices, Renmin University of China, Beijing 100872, China}
\noaffiliation

\date{\today}

\begin{abstract}

So far, experimentally realized quantum anomalous Hall (QAH) insulators all exhibit ferromagnetic order and the QAH effect only occurs at very low temperatures. On the other hand, up to now the QAH effect in antiferromagnetic (AFM) materials has never been reported. In this letter, we realize the QAH effect by proposing a four-band lattice model with static AFM order, which indicates that the QAH effect can be found in AFM materials. Then, as a prototype, we demonstrate that a monolayer CrO can be switched from an AFM Weyl semimetal to an AFM QAH insulator by applying strain, based on symmetry analysis and the first-principles electronic structure calculations. Our work not only proposes a new scenario to search for QAH insulators in materials, but also reveals a way to considerably increase the critical temperature of the QAH phase. 

\end{abstract}

\maketitle
{\it Introduction.} 
 Anomalous Hall effect, which exists in ferromagnetic metals\cite{FerroSemi02, AHE_RMP10}, means that the electric Hall conductance remains to be finite even at zero external magnetic field. Furthermore, if the system becomes insulating, the Hall conductance is quantized to an integer $C$ times $e^2/h$ with $C$ being the Chern number of its band structure. This effect is called the quantum anomalous Hall (QAH) effect.  QAH insulators, i.e. magnetic insulators exhibiting QAH effect ($C\neq0$), as important members in the family of quantum Hall systems, have attracted intensive interest from both theoretical and experimental physicists~\cite{Weng-2015, Liu-2016, Chang-2016, He-2018, NRP-2019, Chang-2022}. In 1988, Haldane proposed a lattice model of spinless fermions to realize integer Hall effect in a staggered magnetic flux with zero net magnetic field\cite{Haldane88}. Later it was proposed that ferromagnetic insulators with spin-orbit coupling (SOC) may exhibit QAH effect \cite{QiWuZhang06, Liu-2016}. And there was also an attempt to derive AFM Chern insulator from the Kane-Mele Hubbard model \cite{Jiang-2018}.  Meanwhile, many materials have been predicted to be QAH insulators\cite{Liu-2008, yu-2010, Wang-2013, Fang-2014, Wu-2014, Ren-2016, Huang-2017, Chen-2017, You-2019, Wu-2019, Zhang-2019, PRR-2019, PRB-2019, PRL-2020, Shi-2021}. So far,  QAH effect has been experimentally observed in four different classes of two-dimensional (2D) systems: thin films of magnetically doped topological insulators\cite{chang-2013}, thin films of the intrinsic magnetic topological insulators\cite{Deng-2020}, moiré materials formed from graphene\cite{Science-2020}, and transition metal dichalcogenides\cite{Nature-2020}. In all of these materials, the QAH plateau only shows up at very low temperatures (of order of 1 Kevin or lower). 

 It was believed that ferromagnetism is necessary for experimental realization of QAH effect\cite{Liu-2016}. However, ferromagnetism can be found easily in metals but rarely in insulators. For this reason, the critical temperature of the observed QAH phase is very low.  On the other hand, there are plenty of AFM insulating materials in nature even above room temperatures. However, so far the QAH effect has never been found in AFM materials, even the corresponding mechanism has never been proposed. Supposing that 
 it can be realized in AFM insulators (especially those containing collinear AFM order), then the realization of QAH effect will be much cheaper. Attractively,  the critical temperature for observing the QAH effect may be greatly increased since the critical temperature of the AFM order can be very high. The resultant AFM QAH insulators can have extensive applications in designing devices. Therefore, the realization of QAH effect in AFM materials is a very important issue both academically and practically.

 \begin{figure*}
\centering
\includegraphics[width=18cm]{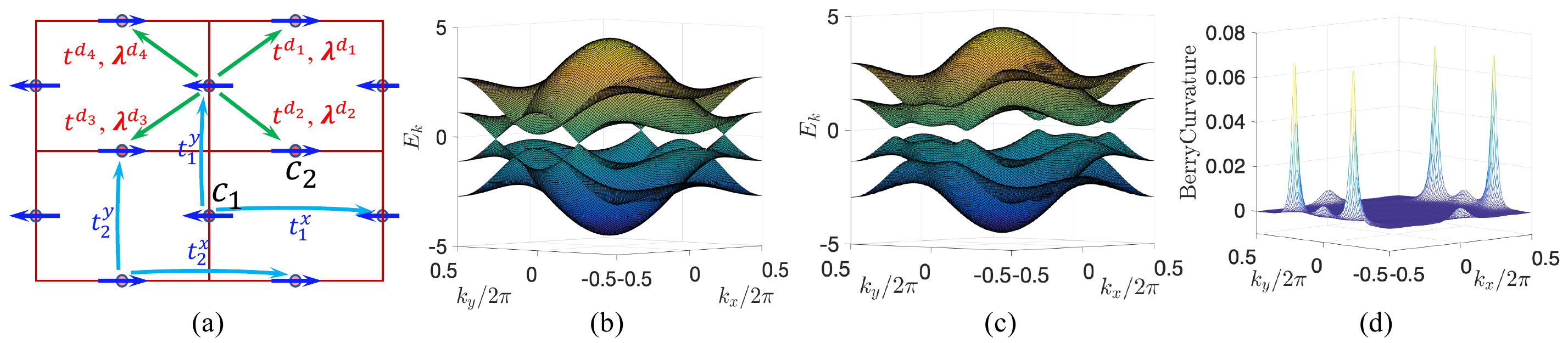}
	\caption{ Model for the AFM QAH insulator. (a) illustration of the Hamiltonian in Eq. (\ref{qahe}); (b) without SOC the ground state being a Weyl semimetal with a pair of Weyl cones on the $(\pi,k_y)$ boundary of the BZ; (c) with SOC the band structure being gapped with Chern number $C=1$; (d) the Berry curvature which is inversion symmetric.}\label{fig:model}
\end{figure*}

Collinear antiferromagnetic (AFM) systems usually possess an `effective time-reversal' symmetry such as $\{\mathcal T| \tau\}$ or $\mathcal I \mathcal T$, where $\mathcal T$ stands for the time reversal operation, $\tau$ denotes the associated fractional translation,  and $\mathcal I$ represents the spatial inversion. Such an `effective time-reversal' symmetry restricts the Chern number of a gapped ground state to be $C=0$, resulting in a trivial band insulator. This limits the study of QAH effect in collinear AFM systems. However, not all collinear AFM systems have the `effective time-reversal' symmetry. Exceptions include the systems having type-I or some of the type-III magnetic space groups. QAH effect can be possibly realized in these collinear AFM insulators.

In this letter, we propose a four-band lattice model to realize QAH effect in a collinear AFM insulator. Furthermore, based on symmetry analysis and the first-principles electronic structure calculations, we illustrate that a monolayer CrO is a collinear AFM semimetal with $C_{2v}$ symmetry protected Weyl cones. And we further show that by applying external strain to lower its point group symmetry, the monolayer CrO can open a gap and be turned into an AFM QAH insulator.

 {\it Lattice model.} We firstly consider a square lattice with two sites (labeled by sublattice index $\alpha=1,2$) in each unit cell which contains local anti-parallel magnetic moments $\pmb m_1=-\pmb m_2 =\pmb m$. The tight-binding model for spin-orbit coupled electrons in the lattice reads
 \beq\label{qahe}
H &=& \sum_{j, d_i}  \Big[ t^{d_i} C_{1,j}^\dag  C_{2, j+ \pmb d_i}  
+  C_{1, j}^\dag  (i  \pmb \lambda^{d_i}\cdot \pmb \sigma) C_{2, j+\pmb d_i} + {\rm h.c.}\Big] \notag \\
      &&+ \sum_{\alpha, j} \Big[ t^x_{\alpha} C_{\alpha , j}^\dag C_{\alpha, j+\pmb x} +  t^{y}_{\alpha}  C_{\alpha, j}^\dag C_{\alpha, j+\pmb y}   + {\rm h.c.} \Big] \notag \\
      &&+ \sum_{\alpha, j} C_{\alpha,j}^\dag \big[  \mu_\alpha\sigma_0 + \pmb m_\alpha\cdot\pmb \sigma\big] C_{\alpha, j},
\eeq
where $C_{\alpha, j}^\dag = (c_{\alpha, j \up}^\dag,c_{\alpha, j\dn}^\dag)$ are electron creation operators, $t^{d_i}$ and $t^x_\alpha, t^y_\alpha$ are the kinetic hopping terms along $\pmb d_{1,2,3,4}$ and $\pmb x, \pmb y$ directions respectively, with $\pmb x=a_1\hat x, \pmb y=a_2\hat y$, $a_{1},a_{2}$ being the lattice constants and $\hat x, \hat y$ being unit vectors, and $\pmb d_{1,3}=\pm{1\over2}(\pmb x+\pmb y), \pmb d_{2,4}=\pm{1\over2}(\pmb x-\pmb y)$. The $i\pmb \lambda^{d_i}\cdot\pmb \sigma$ term denotes spin-orbit coupling, $\pmb m_\alpha = (-1)^\alpha \pmb m$ stands for the static collinear AFM order which couples to the electron spin as a Zeeman field, and $\mu_\alpha$ represents the chemical potential on the $\alpha$-sublattice. We further define $\mu_\pm = {1\over 2}(\mu_1 \pm \mu_2)$, where $\mu_-$ is the staggered potential energy and $\mu_+$ is adopted such that the energy bands are half-filled. Here we ignore the correlation between the electrons by simply assuming that the AFM moment couples to the electron spin as a Zeeman field. The effect of electron correlations,  such as the Anderson or the Kondo coupling terms, will be left for future study.

Since inversion symmetry is crucial to guarantee a nonzero Chern number, it requires that $t^{d_1}=t^{d_3}$, $t^{d_2}=t^{d_4}$  and $\pmb \lambda^{d_1}=\pmb \lambda^{d_3}$, $\pmb \lambda^{d_2}=\pmb \lambda^{d_4}$.  For simplicity, we set $t^{d_1}=t^{d_2}=t^d$ and $\pmb \lambda^{d_1} = -\pmb \lambda^{d_2} = \pmb \lambda $. Furthermore, to ensure that the energy bands have a full gap, we let $t^{x,y}_\alpha =(-1)^\alpha t^{x,y}$. 
After the Fourier transformation, the above Hamiltonian can be rewritten in momentum space as 
\beq\label{Hk}
H = \sum_{k} \big( C_{1k}^\dag \Gamma^{12}_k  C_{2k} + {\rm h.c.}\big) + \sum_{\alpha, k} C_{\alpha k}^\dag \Gamma^{(\alpha)}_{k} C_{\alpha k},
\eeq
with $C_{\alpha k}^\dag =(c_{\alpha k\up}^\dag,c_{\alpha k\dn}^\dag)$ the fermion operators, $\Gamma^{12}_k = 2(t^{d} +i\pmb\lambda \cdot\pmb\sigma)\cos({k_x\over2}+{k_y\over2}) + 2(t^{d} - i\pmb\lambda \cdot\pmb\sigma)\cos({k_x\over2}-{k_y\over2}) $ the inter-sublattice terms and $\Gamma^{(\alpha)}_{k}=(-1)^\alpha 2(t^x \cos{k_x} +t^y \cos{k_y}) + (\mu_\alpha + \pmb m_\alpha\cdot\pmb \sigma)$ the intra-sublattice terms. 

We firstly turn off the SOC by setting $\pmb\lambda^{d_1}=\pmb\lambda^{d_2}=0$, then the system has a spin point group symmetry\cite{Litvin1974,QHLiuXGWanprx,YangLiuFang21,GWLLL21} generated by $(E||C_2^z), (E||C_2^x), (E||\mathcal I)$ and $(C_2^\perp \mathcal T||\mathcal T), (C_2^{\pmb m}||E)$, where the notation $(g||h)$ denotes a combined operation of the spin operation $g$ and the lattice operation $h$, and $C_2^\perp$/$C_2^{\pmb m}$ respectively stand for the  2-fold spin rotation along the axis perpendicular/parallel to the $\pmb m$-direction. In this case the $\Gamma ^{12}_k$ term vanishes on the boundary of the BZ. In other words, the two species of fermions $C_{1k}$ and $C_{2k}$ do not hybridize if $k_x=\pi$ or $k_y=\pi$. It can be shown that on the 
boundary line $(k_x, \pi)$, the $C_{1k}, C_{2k}$ bands carry quantum numbers $1$ and $-1$ of the symmetry operation $(E||C_2^x)$ [or $(E||M_{y})$], respectively.  Similarly, on the 
line $(\pi,k_y)$ the $C_{1k}, C_{2k}$ bands respectively carry quantum numbers $-1$ and $1$ of the symmetry operation $(E||C_2^y)$ [or $(E||M_{x})$] . Owing to these different quantum numbers, if the $C_{1k}$ and $C_{2k}$ bands are inverted, the band crossing will be protected from opening a gap. Resultantly, on the boundary lines $(k_x, \pi)$ and/or $(\pi,k_y)$ there will be pairs of Weyl cones (see Fig.\ref{fig:model}(b) as an example) among which the two in each pair are related by the inversion $\mathcal I$ operation. It can be further shown that when the condition (\ref{mu_qahe}) given below is satisfied, there will be odd pairs of Weyl cones on the BZ boundary [see the Supplemental Materials \cite{sm}(SM)]. In the SM we also show that the Weyl cones are robust under perturbations as long as the symmetry $(E||\mathcal I)\times(C_2^\perp \mathcal T||\mathcal T) = (C_2^\perp T||\mathcal {IT})$ is unbroken.

Then we turn on the SOC. In this case, the physical symmetry is described by a magnetic point group owing to the locking of the lattice rotation $g$ and the spin rotation $g$ (we will simply denote $(g||g)$ as $g$ in the following). Generally, if $\pmb \lambda\cdot\pmb m\neq0$ then all the Weyl cones obtain a mass and the energy band is fully gapped (see Fig.\ref{fig:model}(c)). Noticing that the SOC does not break the inversion symmetry $\mathcal I$, and that the pattern of the Berry curvature of the occupied bands is invariant under $\mathcal I$  (see Fig.\ref{fig:model}(d)), each pair of Weyl cones (related by $\mathcal I$) together contribute either $1$ or $-1$ to the total Chern number (recalling that a single cone carries a $\pi$ Berry phase or $\pm1/2$ Chern number). Therefore, if there are an odd number of pairs of Weyl cones, the Chern number in the gapped state must be nonzero and the system will exhibit QAH effect. 


\begin{figure}
	\includegraphics[width=6.cm]{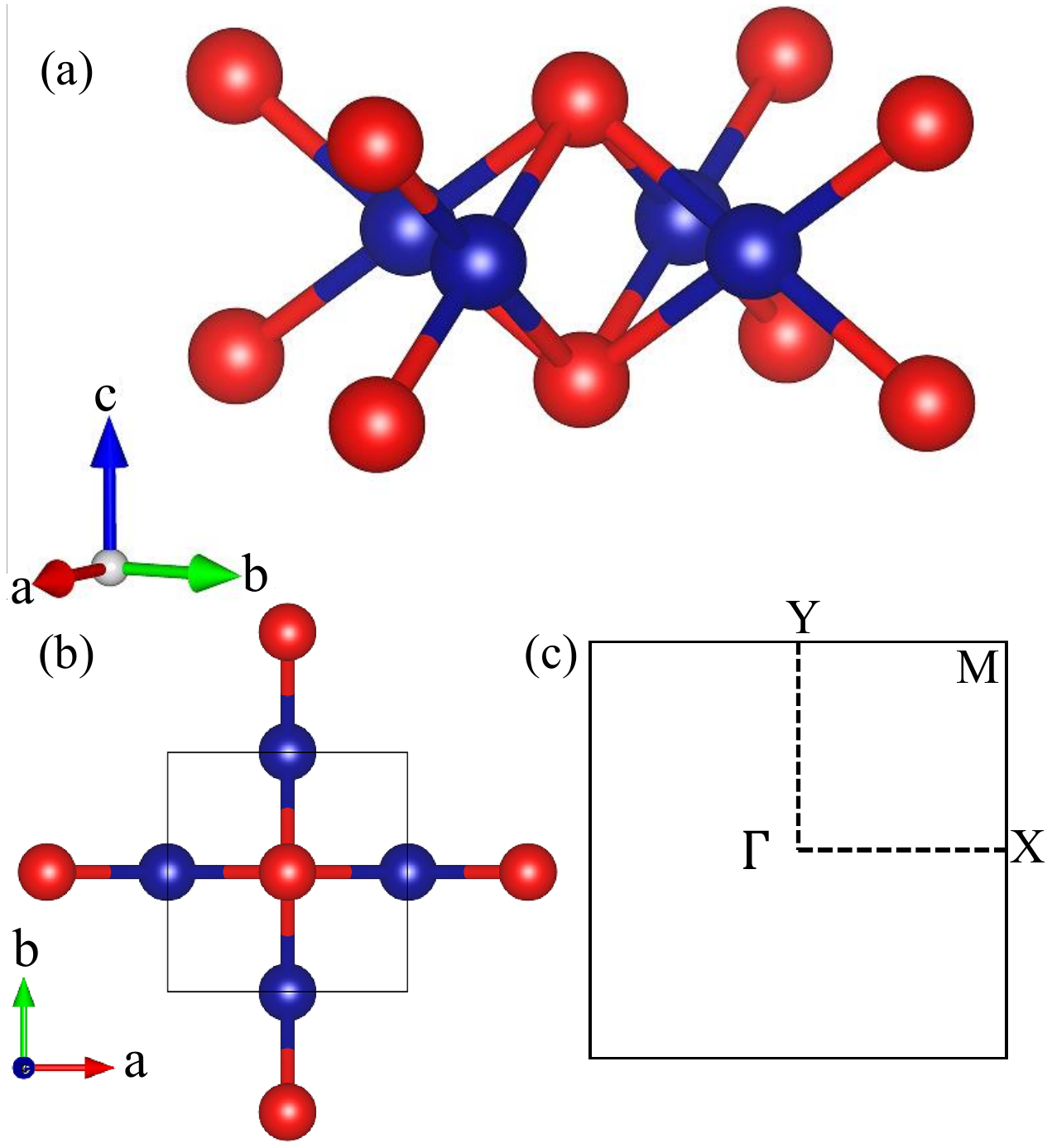}
	\caption{Crystal structure of the monolayer CrO viewed along (a) [110] and (b) [001] directions. (c) The Brillouin zone  and the corresponding high-symmetry points. The red and blue balls represent the O and Cr atoms, respectively.}\label{fig:1}
\end{figure}

The conditions for the nontrivial total Chern number ($\pm1$) can be summarized as the following inequalities,
\beq
&t^d\neq0,\ \ \pmb\lambda \cdot \pmb m\neq 0, &\label{m_qahe}
\\&\Big||\pmb m|-2|t^y-t^x|\Big|< |\mu_-| < |\pmb m|+2|t^y-t^x|.&\label{mu_qahe}
\eeq
These conditions restrict the crystalline symmetry of the potential materials (see SM for detailed discussions). If the AFM order lies in the lattice plane(without lose of generality we assume $\pmb m\parallel \hat y$), then the QAH effect can be realized in triclinic lattice with symmetry $\bar 1=\{E,\mathcal I\}$ or monoclinic lattice with magnetic point group $2'/m'=\{E,\mathcal I, C_2^x\mathcal T, M_x\mathcal T\}$. On the other hand, if the AFM order is perpendicular to the lattice plane (namely $\pmb m\parallel\hat z$), then the conditions (\ref{m_qahe}) and (\ref{mu_qahe}) can be satisfied in orthogonal lattice with magnetic point group symmetry $m'm'm=\{E, C_2^x\mathcal T, C_2^y\mathcal T, C_2^z, \mathcal I, M_x\mathcal T,  M_y\mathcal T, M_z\}$.

If $\pmb\lambda \cdot\pmb m=0$, and if $\pmb m$ lies in the lattice plane, e.g. $\pmb m\parallel y$, then the $(C_2^y||C_2^y)$ or the $(C_2^y||M_y)=(C_2^y||C_2^y)\times(E||\mathcal I)$ symmetry can protect the Weyl cones on the BZ boundary from being gapped out (see SM). This means that if the AFM order is parallel to an in-plane $C_2$ axis, then the QAH effect cannot be realized through our mechanism.

Now we stress that the spin-orbit coupling term in model (\ref{qahe}) is inversion symmetric. When $\pmb \lambda^{d_1} = \lambda\hat x, \pmb \lambda^{d_2} = \lambda\hat y$, it looks like a Rashba SOC term, but the difference is that the Rashba term breaks $\mathcal I$ symmetry while our model (\ref{qahe}) does not. In this case, the QAH effect can be realized if conditions $\pmb m\cdot(\hat x+\hat y)\neq 0$ and (\ref{mu_qahe}) are satisfied.

It is interesting to compare the model (\ref{qahe}) with the Haldane model\cite{Haldane88} and the FM QAH model\cite{QiWuZhang06}. Without introducing external magnetic fields, all of these models have a gapped ground state with a nonzero Chern number. In the Haldane model, a staggered flux is needed in which the fermions obtain the Aharonov-Bohm phase. Although the total flux is zero, a staggered flux is not easy to implement in practice. In the FM QAH model, the staggered flux is replaced by the Berry phase resulting from spin-orbit coupling, and the role of the FM order is to polarize the spin of the electrons. However, the FM critical temperature in insulators is generally very low. In our model, the AFM order provides a local Zeeman field for the electrons, the SOC and the sub-lattice dependent chemical potentials guarantee a full gap and a nontrivial Chern number. The advantage is that the  AFM critical temperature can be very high (even with an order of 1000 Kelvin). Therefore, if the band gap is not too small, then hopefully the QAH effect can be realized in experiments close to room temperatures.

As discussed above, the low-energy effective model (\ref{qahe})  reveals the possibility that QAH effect can exist in AFM insulators. In the following, by studying a concrete material as a prototype --- a monolayer CrO, we illustrate that QAH effect can indeed be obtained in AFM materials.

{\it Candidate Material.} A monolayer CrO has a square lattice structure whose symmetry is described by the symmorphic space group P4-mmm. The point group of P4-mmm is D$_{4h}$ generated by $C_{4}^{z}, C_{2}^{x}$ and $ \mathcal I$. The unit cell of a monolayer CrO contains two O atoms plus two Cr atoms (see Fig.\ref{fig:1} (a) and (b)) and the corresponding BZ is shown in Fig.\ref{fig:1} (c). The Cr$^{2+}$ ions have a fairly strong correlation and they interact with each other via super-exchange interactions mediated by the $O$ atoms. It was suggested in Ref.\cite{Chen-2021} that monolayer CrO is in antiferromagnetic phase with a high Neel temperature.

\begin{figure}
	\includegraphics[width=8cm]{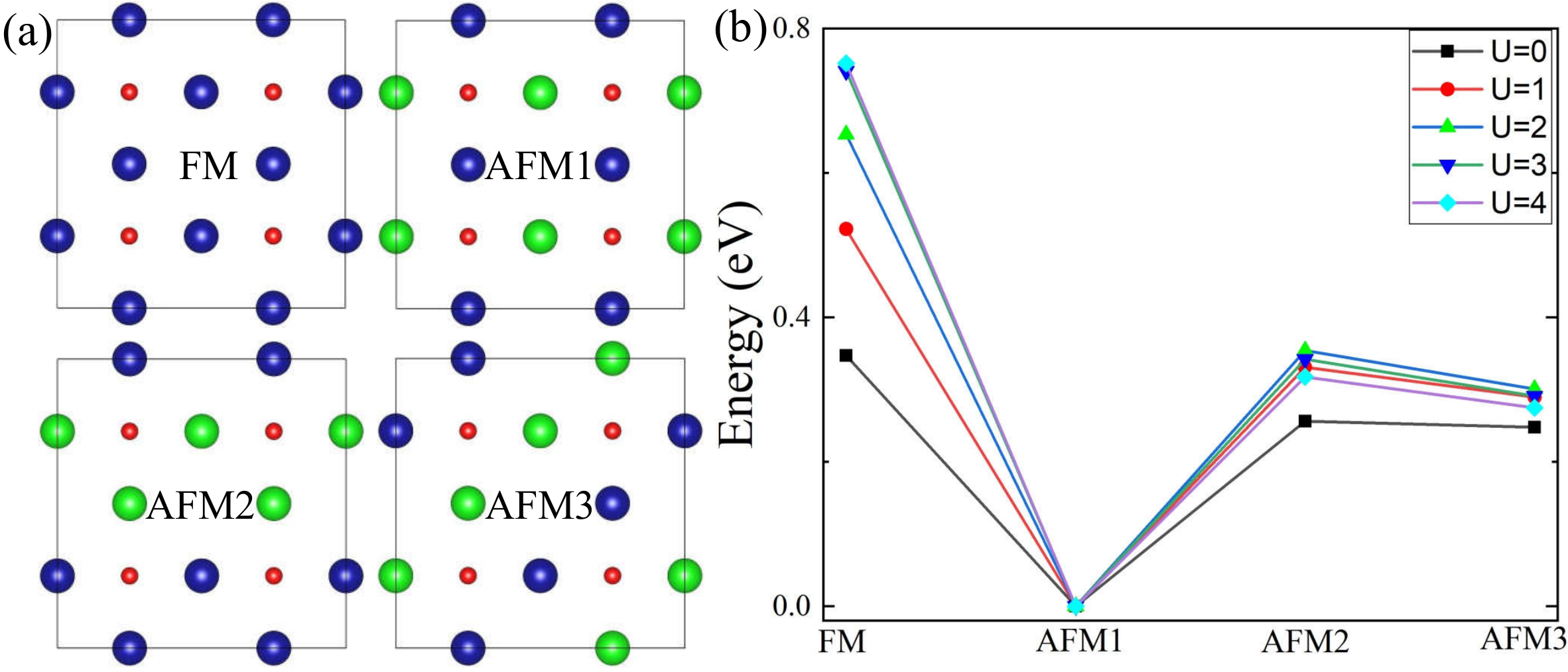}
	\caption{(a) Cartoon pictures for four possible magnetic structures of a monolayer CrO, where  FM and AFM represent ferromagnetism and antiferromagnetism, respectively. The blue and green balls represent Cr atoms with opposite spin. (b) The relative energies (per Cr atom) of the four ordered states under different Hubbard U (eV).}\label{fig:2}
\end{figure}

To confirm the magnetic structure of a monolayer CrO, we compare the energies of four trial states with different magnetic orders under different Hubbard $U$. As shown in Fig.\ref{fig:2}(a), indeed the AFM1 state (i.e. the collinear Neel state) is always  lowest in energy. Notice that the AFM1 order has no `effective time-reversal symmetry' like $\mathcal I \mathcal T$ or $\{\mathcal T|\tau\}$. Furthermore, since the electronic band structure is sensitive to the Hubbard $U$, we need to determine the proper value of $U$. By comparing the electronic band structure of Heyd-Scuseria-Ernzerhof (HSE) hybrid functional within the framework of HSE06\cite{HSE-2006}, we find that $U=3.2$ eV, which is slightly less than 3.5 eV as proposed in \cite{Chen-2021}. This is due to the fact that our relaxed lattice parameter is 1.4 percent less than that adopted in \cite{Chen-2021}. 

\begin{figure}
	\includegraphics[width=8 cm]{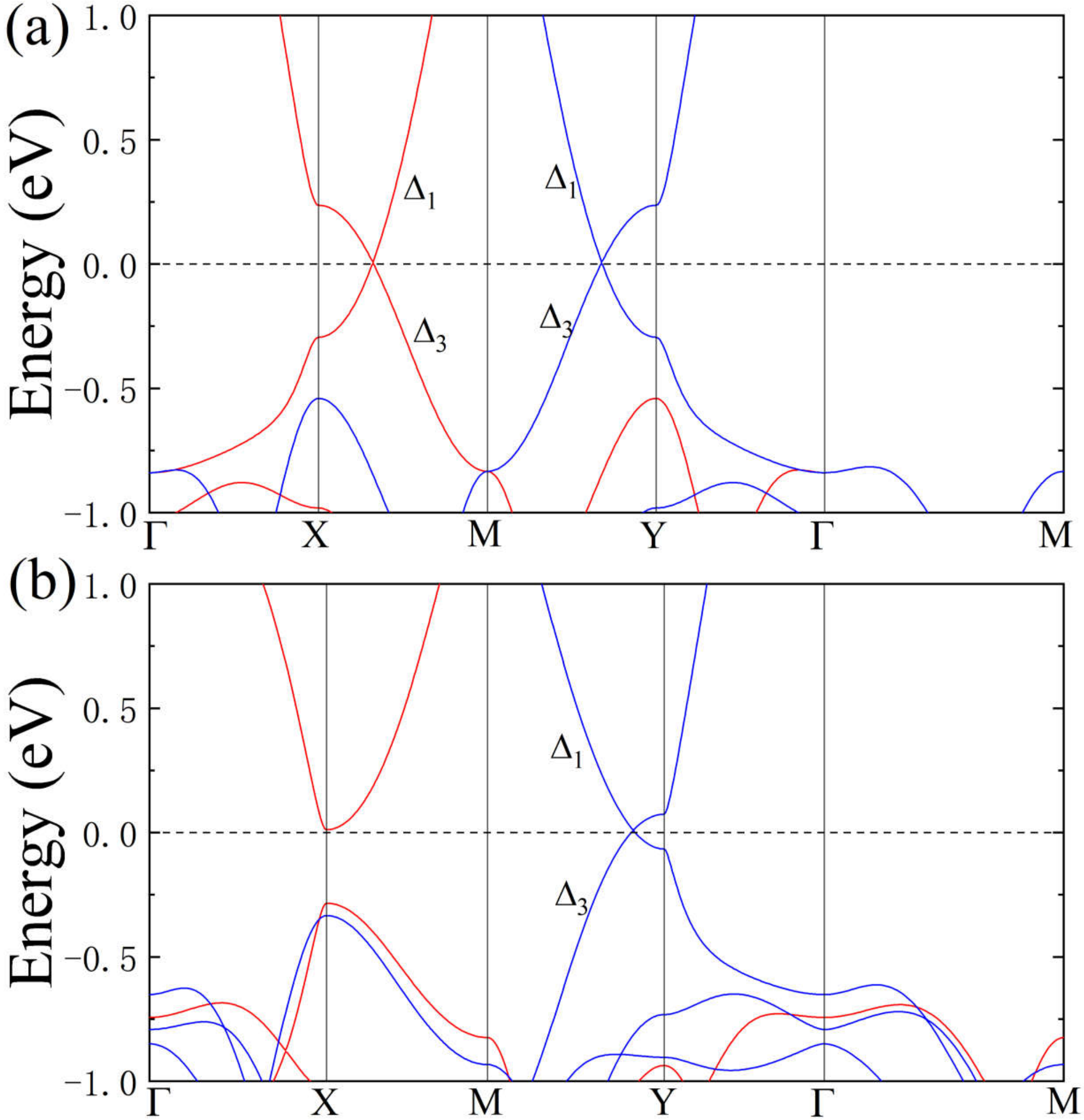}
	\caption{ The band structure of monolayer CrO along the high-symmetry directions, (a) no SOC, (b) no SOC, but tensile strain 10 percent along b direction. The spin-up and spin-down bands in (a) and (b) are marked in red and blue, respectively. The $\Delta_{1}$ and $\Delta_{3}$ represent different irreducible representations of the little co-group $C_{2v}$.}\label{fig:3}
\end{figure}
 
When ignoring the SOC, a monolayer CrO has a spin point group symmetry whose generators are $(\mathcal T||\mathcal TC_4^z), (E||M_z), (E||M_x),(\mathcal T ||\mathcal T M_{x+y})$ and $(C_2^\perp \mathcal T||\mathcal T),(C_2^{\pmb m}||E)$. According to the $(\mathcal T||\mathcal TC_{2}^{x\pm y})=(E||M_z)(\mathcal T ||\mathcal T M_{x\mp y})$ symmetry, the spin-up and spin-down are degenerate on the high-symmetry lines $\Gamma$-$M$ along the $x+y$ or $x-y$ direction. Away from these two high symmetry lines, the spin degeneracy is generally lifted as illustrated in Fig.\ref{fig:3}(a), where the red and blue lines represent spin-up and spin-down bands, respectively. Furthermore,  the band inversion occurs around the X and Y points which are related to each other by the $(\mathcal T||\mathcal TC_4^z)$ symmetry and hence carry opposite spins. Since the two inverted bands carry different quantum numbers of $(E||C_2^x)$ or $(E||C_2^y)$, the band inversion results in a band crossing with two pairs of Weyl points\cite{Chen-2021}. The two pairs of Weyl cones around the X,Y points can also appear in the model (\ref{qahe}) when $t^x=t^y$ and $\pmb\lambda=0$.

If the $(\mathcal T||\mathcal TC_{4}^{z})$ symmetry is preserved, then after the two pairs of Weyl cones being gapped out, the Berry curvature contributed from these cones will exactly cancel each other, resulting in a trivial Chern number. Therefore, in order to realize the QAH effect in a monolayer CrO, the $(\mathcal T||\mathcal TC_{4}^{z})$ symmetry must be removed. To this end, we apply a tensile strain 10 percent along the b-direction. Then the crystalline point group of the strained monolayer CrO reduces to $D_{2h}$ which is generated by $C_2^z, C_2^x$ and $\mathcal  I$.

\begin{figure}
	{\includegraphics[width=8.7cm]{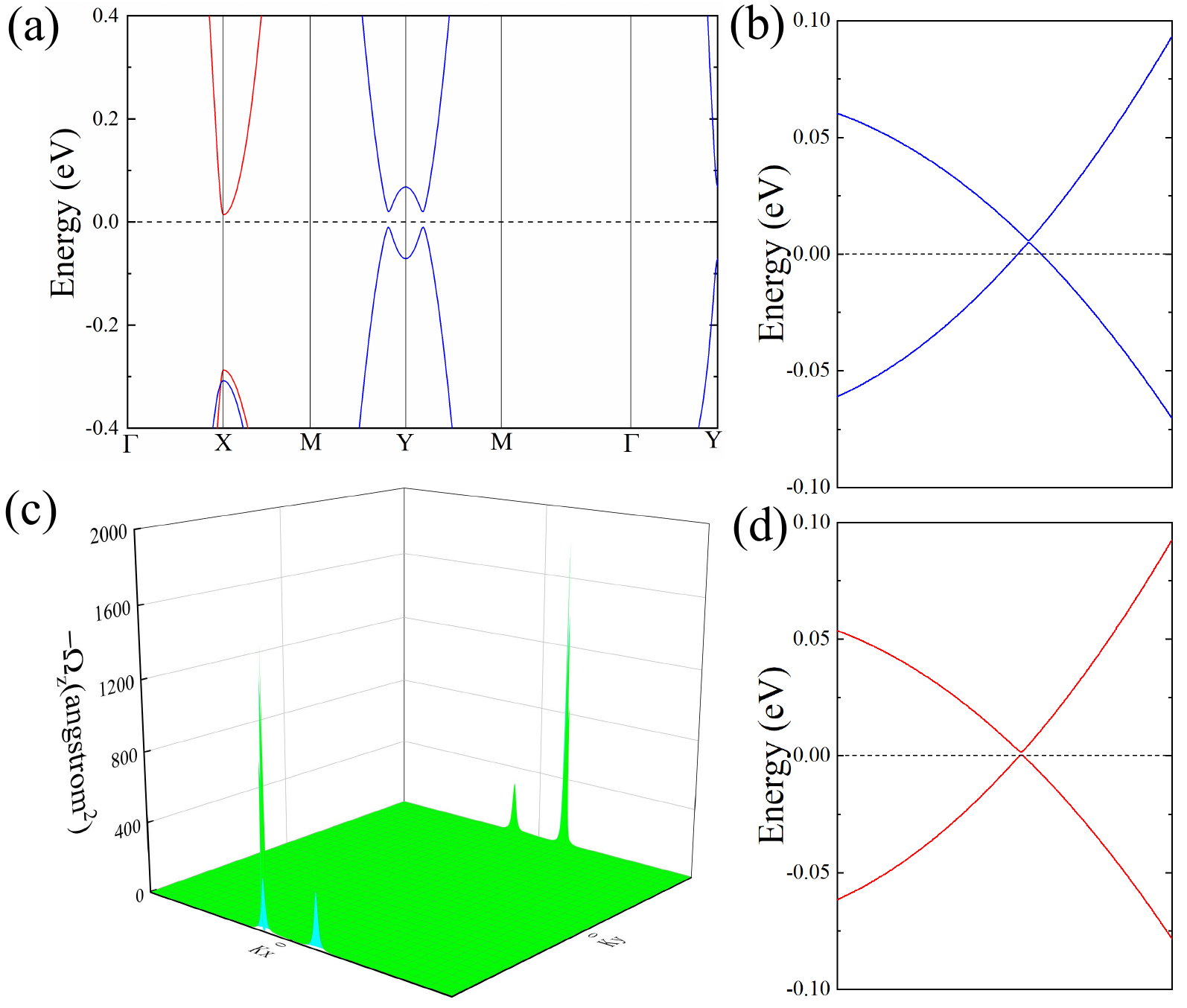}} 
	\caption{ The Weyl cones of a monolayer CrO shift their positions from the M-Y high symmetry line to the inside of the BZ by the tensile and shear strain, and are further gaped out if SOC is turned on. (a) \& (b) show the electronic band structure (with strain and without SOC) along the M-Y line and 
	the line $k_y =0.49747\times{2\pi\over a_2}$ respectively, it can be seen that the Weyl points shift toward the inside of the BZ. (c) \& (d) respectively show the Berry curvature and the dispersion (around the Weyl point) of the gapped  bands when SOC is turned on. }\label{fig:4}
\end{figure}

Under the above tensile strain, the calculated electronic band structure is shown in Fig.\ref{fig:3}(b), where the pair of Weyl points on the X-M axis merge and gap out with the disappearing of band inversion. But the pair of Weyl points on the Y-M axis remain stable. When considering the SOC, our calculations indicate that the magnetic moments are along the $\hat y$-direction in the lattice plane. Then the monolayer CrO has a magnetic point group symmetry $m'mm'=\{E, C_2^x\mathcal T, C_2^y, C_2^z\mathcal T, \mathcal I, M_x\mathcal T,  M_y, M_z\mathcal T\}$ 
However, due to the $C_2^z\mathcal T\equiv (C_2^z\mathcal T||C_2^z\mathcal T)$ symmetry, the Berry curvature is always zero in the whole BZ and the Weyl cones are still robust. Therefore, the $C_2^z\mathcal T$ symmetry has to be removed to gap out the Weyl cones and then to realize the QAH effect.



To break the  $C_2^z\mathcal T$ symmetry, we further apply shear strain (up to 3 percent) such that the two out-of-plane O atoms obtain a slight displacement along the in-plane diagonal direction (the inversion symmetry is always kept). At this time, the monolayer CrO has only space-inversion symmetry $\mathcal I$. Finally, the pair of Weyl cones open a gap for amount $\sim$1 meV  (see Fig.\ref{fig:4}(d) for the gapped cone) and the resultant gapped ground state has a Chern number $C=-1$, which makes the strained monolayer CrO a QAH insulator. As shown in Fig.\ref{fig:4}(c), the Berry curvature mainly concentrates around the gapped Weyl points.
  
It should be emphasized that SOC plays an important role in giving mass to the Weyl cones. If SOC is absent, the  Weyl cones are robust against the strains [see Fig.\ref{fig:4}(a)\&(b) where the Weyl points on the Y-M line shift to $\left( 0.09025, 0.49747\right)$ and  $\left( -0.09025, -0.49747\right)$] owing to the protection of the  spin point group symmetry element $(C_2^z\mathcal T||\mathcal {IT})$.  So it is not a wonder that the size of the gap is rather small since the SOC in CrO is very weak.  For this reason, the QAH effect can only  be observed at about 10 Kelvin even though the Neel temperature of a monolayer CrO is estimated to be above the room temperature\cite{Chen-2021}.
 
As mentioned, the monolayer CrO is just a prototype to demonstrate the mechanism of realizing QAH effect in an AFM system. The critical temperature of QAH effect can be promisingly enhanced in other candidate materials with stronger SOC. We believe that our work will arouse further interest in the study of QAH effect in AFM systems.


In summary, we come up with a four-band lattice model to illustrate that QAH effect can exist in antiferromagnetism. Furthermore, we predict that the monolayer CrO is a candidate AFM material to realize the QAH effect. Based on symmetry analysis and the first-principles electronic structure calculations, we first show that a monolayer CrO is an AFM semimetal containing Weyl cones. Then we demonstrate that by adding tensile and shear stain, the monolayer CrO can be turned into an AFM QAH insulator. Our study reveals a new scenario to realize QAH insulator in AFM materials. Especially, once QAH insulator is realized in AFM systems with strong SOC, it may allow people to observe QAH effect close to room temperatures.

{\it Acknowledgments.} We thank J.-W. Liu, C.-C. Liu,  X.-H. Kong, Z.-M. Yu for valueable discussions. This work was financially supported by the NSF of China (No. 12134020, No. 11974421 and No. 11934020). 



%


\begin{thebibliography}{11}%
\makeatletter
\providecommand \@ifxundefined [1]{%
 \@ifx{#1\undefined}
}%
\providecommand \@ifnum [1]{%
 \ifnum #1\expandafter \@firstoftwo
 \else \expandafter \@secondoftwo
 \fi
}%
\providecommand \@ifx [1]{%
 \ifx #1\expandafter \@firstoftwo
 \else \expandafter \@secondoftwo
 \fi
}%
\providecommand \natexlab [1]{#1}%
\providecommand \enquote  [1]{``#1''}%
\providecommand \bibnamefont  [1]{#1}%
\providecommand \bibfnamefont [1]{#1}%
\providecommand \citenamefont [1]{#1}%
\providecommand \href@noop [0]{\@secondoftwo}%
\providecommand \href [0]{\begingroup \@sanitize@url \@href}%
\providecommand \@href[1]{\@@startlink{#1}\@@href}%
\providecommand \@@href[1]{\endgroup#1\@@endlink}%
\providecommand \@sanitize@url [0]{\catcode `\\12\catcode `\$12\catcode
  `\&12\catcode `\#12\catcode `\^12\catcode `\_12\catcode `\%12\relax}%
\providecommand \@@startlink[1]{}%
\providecommand \@@endlink[0]{}%
\providecommand \url  [0]{\begingroup\@sanitize@url \@url }%
\providecommand \@url [1]{\endgroup\@href {#1}{\urlprefix }}%
\providecommand \urlprefix  [0]{URL }%
\providecommand \Eprint [0]{\href }%
\providecommand \doibase [0]{https://doi.org/}%
\providecommand \selectlanguage [0]{\@gobble}%
\providecommand \bibinfo  [0]{\@secondoftwo}%
\providecommand \bibfield  [0]{\@secondoftwo}%
\providecommand \translation [1]{[#1]}%
\providecommand \BibitemOpen [0]{}%
\providecommand \bibitemStop [0]{}%
\providecommand \bibitemNoStop [0]{.\EOS\space}%
\providecommand \EOS [0]{\spacefactor3000\relax}%
\providecommand \BibitemShut  [1]{\csname bibitem#1\endcsname}%
\let\auto@bib@innerbib\@empty
\bibitem [{\citenamefont {Bouhon}\ \emph {et~al.}(2021)\citenamefont {Bouhon},
  \citenamefont {Lange},\ and\ \citenamefont {Slager}}]{Bouhon-2021PRB}%
  \BibitemOpen
  \bibfield  {author} {\bibinfo {author} {\bibfnamefont {A.}~\bibnamefont
  {Bouhon}}, \bibinfo {author} {\bibfnamefont {G.~F.}\ \bibnamefont {Lange}},\
  and\ \bibinfo {author} {\bibfnamefont {R.-J.}\ \bibnamefont {Slager}},\
  }\bibfield  {title} {\bibinfo {title} {Topological correspondence between
  magnetic space group representations and subdimensions},\ }\href
  {https://doi.org/10.1103/PhysRevB.103.245127} {\bibfield  {journal} {\bibinfo
   {journal} {Phys. Rev. B}\ }\textbf {\bibinfo {volume} {103}},\ \bibinfo
  {pages} {245127} (\bibinfo {year} {2021})}\BibitemShut {NoStop}%
\bibitem [{\citenamefont {Jiang}\ \emph {et~al.}(2018)\citenamefont {Jiang},
  \citenamefont {Zhou}, \citenamefont {Dai},\ and\ \citenamefont
  {Wang}}]{Jiang-2018}%
  \BibitemOpen
  \bibfield  {author} {\bibinfo {author} {\bibfnamefont {K.}~\bibnamefont
  {Jiang}}, \bibinfo {author} {\bibfnamefont {S.}~\bibnamefont {Zhou}},
  \bibinfo {author} {\bibfnamefont {X.}~\bibnamefont {Dai}},\ and\ \bibinfo
  {author} {\bibfnamefont {Z.}~\bibnamefont {Wang}},\ }\bibfield  {title}
  {\bibinfo {title} {Antiferromagnetic chern insulators in noncentrosymmetric
  systems},\ }\href {https://doi.org/10.1103/PhysRevLett.120.157205} {\bibfield
   {journal} {\bibinfo  {journal} {Phys. Rev. Lett.}\ }\textbf {\bibinfo
  {volume} {120}},\ \bibinfo {pages} {157205} (\bibinfo {year}
  {2018})}\BibitemShut {NoStop}%
\bibitem [{\citenamefont {Bl\"ochl}(1994)}]{PAW-1994}%
  \BibitemOpen
  \bibfield  {author} {\bibinfo {author} {\bibfnamefont {P.~E.}\ \bibnamefont
  {Bl\"ochl}},\ }\bibfield  {title} {\bibinfo {title} {Projector augmented-wave
  method},\ }\href {https://doi.org/10.1103/PhysRevB.50.17953} {\bibfield
  {journal} {\bibinfo  {journal} {Phys. Rev. B}\ }\textbf {\bibinfo {volume}
  {50}},\ \bibinfo {pages} {17953} (\bibinfo {year} {1994})}\BibitemShut
  {NoStop}%
\bibitem [{\citenamefont {Kresse}\ and\ \citenamefont
  {Joubert}(1999)}]{PAW-1999}%
  \BibitemOpen
  \bibfield  {author} {\bibinfo {author} {\bibfnamefont {G.}~\bibnamefont
  {Kresse}}\ and\ \bibinfo {author} {\bibfnamefont {D.}~\bibnamefont
  {Joubert}},\ }\bibfield  {title} {\bibinfo {title} {From ultrasoft
  pseudopotentials to the projector augmented-wave method},\ }\href
  {https://doi.org/10.1103/PhysRevB.59.1758} {\bibfield  {journal} {\bibinfo
  {journal} {Phys. Rev. B}\ }\textbf {\bibinfo {volume} {59}},\ \bibinfo
  {pages} {1758} (\bibinfo {year} {1999})}\BibitemShut {NoStop}%
\bibitem [{\citenamefont {Kresse}\ and\ \citenamefont
  {Furthmüller}(1996)}]{Cms-1996}%
  \BibitemOpen
  \bibfield  {author} {\bibinfo {author} {\bibfnamefont {G.}~\bibnamefont
  {Kresse}}\ and\ \bibinfo {author} {\bibfnamefont {J.}~\bibnamefont
  {Furthmüller}},\ }\bibfield  {title} {\bibinfo {title} {Efficiency of
  ab-initio total energy calculations for metals and semiconductors using a
  plane-wave basis set},\ }\href
  {https://doi.org/https://doi.org/10.1016/0927-0256(96)00008-0} {\bibfield
  {journal} {\bibinfo  {journal} {Computational Materials Science}\ }\textbf
  {\bibinfo {volume} {6}},\ \bibinfo {pages} {15} (\bibinfo {year}
  {1996})}\BibitemShut {NoStop}%
\bibitem [{\citenamefont {Kresse}\ and\ \citenamefont
  {Furthm\"uller}(1996)}]{tnc-1996}%
  \BibitemOpen
  \bibfield  {author} {\bibinfo {author} {\bibfnamefont {G.}~\bibnamefont
  {Kresse}}\ and\ \bibinfo {author} {\bibfnamefont {J.}~\bibnamefont
  {Furthm\"uller}},\ }\bibfield  {title} {\bibinfo {title} {Efficient iterative
  schemes for ab initio total-energy calculations using a plane-wave basis
  set},\ }\href {https://doi.org/10.1103/PhysRevB.54.11169} {\bibfield
  {journal} {\bibinfo  {journal} {Phys. Rev. B}\ }\textbf {\bibinfo {volume}
  {54}},\ \bibinfo {pages} {11169} (\bibinfo {year} {1996})}\BibitemShut
  {NoStop}%
\bibitem [{\citenamefont {Perdew}\ \emph {et~al.}(1996)\citenamefont {Perdew},
  \citenamefont {Burke},\ and\ \citenamefont {Ernzerhof}}]{GGA-1996}%
  \BibitemOpen
  \bibfield  {author} {\bibinfo {author} {\bibfnamefont {J.~P.}\ \bibnamefont
  {Perdew}}, \bibinfo {author} {\bibfnamefont {K.}~\bibnamefont {Burke}},\ and\
  \bibinfo {author} {\bibfnamefont {M.}~\bibnamefont {Ernzerhof}},\ }\bibfield
  {title} {\bibinfo {title} {Generalized gradient approximation made simple},\
  }\href {https://doi.org/10.1103/PhysRevLett.77.3865} {\bibfield  {journal}
  {\bibinfo  {journal} {Phys. Rev. Lett.}\ }\textbf {\bibinfo {volume} {77}},\
  \bibinfo {pages} {3865} (\bibinfo {year} {1996})}\BibitemShut {NoStop}%
\bibitem [{\citenamefont {Chen}\ \emph {et~al.}(2021)\citenamefont {Chen},
  \citenamefont {Wang}, \citenamefont {Li},\ and\ \citenamefont
  {Sanyal}}]{Chen-2021}%
  \BibitemOpen
  \bibfield  {author} {\bibinfo {author} {\bibfnamefont {X.}~\bibnamefont
  {Chen}}, \bibinfo {author} {\bibfnamefont {D.}~\bibnamefont {Wang}}, \bibinfo
  {author} {\bibfnamefont {L.}~\bibnamefont {Li}},\ and\ \bibinfo {author}
  {\bibfnamefont {B.~S.}\ \bibnamefont {Sanyal}},\ }\href
  {https://doi.org/arxiv.org/abs/2104.07390} {\bibfield  {journal} {\bibinfo
  {journal} {ArXiv:}\ }\textbf {\bibinfo {volume} {2104}},\ \bibinfo {pages}
  {07390} (\bibinfo {year} {2021})}\BibitemShut {NoStop}%
\bibitem [{\citenamefont {Marzari}\ and\ \citenamefont
  {Vanderbilt}(1997)}]{wannier-1997}%
  \BibitemOpen
  \bibfield  {author} {\bibinfo {author} {\bibfnamefont {N.}~\bibnamefont
  {Marzari}}\ and\ \bibinfo {author} {\bibfnamefont {D.}~\bibnamefont
  {Vanderbilt}},\ }\bibfield  {title} {\bibinfo {title} {Maximally localized
  generalized wannier functions for composite energy bands},\ }\href
  {https://doi.org/10.1103/PhysRevB.56.12847} {\bibfield  {journal} {\bibinfo
  {journal} {Phys. Rev. B}\ }\textbf {\bibinfo {volume} {56}},\ \bibinfo
  {pages} {12847} (\bibinfo {year} {1997})}\BibitemShut {NoStop}%
\bibitem [{\citenamefont {Souza}\ \emph {et~al.}(2001)\citenamefont {Souza},
  \citenamefont {Marzari},\ and\ \citenamefont {Vanderbilt}}]{wannier-2001}%
  \BibitemOpen
  \bibfield  {author} {\bibinfo {author} {\bibfnamefont {I.}~\bibnamefont
  {Souza}}, \bibinfo {author} {\bibfnamefont {N.}~\bibnamefont {Marzari}},\
  and\ \bibinfo {author} {\bibfnamefont {D.}~\bibnamefont {Vanderbilt}},\
  }\bibfield  {title} {\bibinfo {title} {Maximally localized wannier functions
  for entangled energy bands},\ }\href
  {https://doi.org/10.1103/PhysRevB.65.035109} {\bibfield  {journal} {\bibinfo
  {journal} {Phys. Rev. B}\ }\textbf {\bibinfo {volume} {65}},\ \bibinfo
  {pages} {035109} (\bibinfo {year} {2001})}\BibitemShut {NoStop}%
\bibitem [{\citenamefont {Wu}\ \emph {et~al.}(2018)\citenamefont {Wu},
  \citenamefont {Zhang}, \citenamefont {Song}, \citenamefont {Troyer},\ and\
  \citenamefont {Soluyanov}}]{Wu-2018}%
  \BibitemOpen
  \bibfield  {author} {\bibinfo {author} {\bibfnamefont {Q.}~\bibnamefont
  {Wu}}, \bibinfo {author} {\bibfnamefont {S.}~\bibnamefont {Zhang}}, \bibinfo
  {author} {\bibfnamefont {H.-F.}\ \bibnamefont {Song}}, \bibinfo {author}
  {\bibfnamefont {M.}~\bibnamefont {Troyer}},\ and\ \bibinfo {author}
  {\bibfnamefont {A.~A.}\ \bibnamefont {Soluyanov}},\ }\bibfield  {title}
  {\bibinfo {title} {Wanniertools: An open-source software package for novel
  topological materials},\ }\href
  {https://doi.org/https://doi.org/10.1016/j.cpc.2017.09.033} {\bibfield
  {journal} {\bibinfo  {journal} {Computer Physics Communications}\ }\textbf
  {\bibinfo {volume} {224}},\ \bibinfo {pages} {405} (\bibinfo {year}
  {2018})}\BibitemShut {NoStop}%
\end{thebibliography}%


\begin{thebibliography}{40}%
\makeatletter
\providecommand \@ifxundefined [1]{%
 \@ifx{#1\undefined}
}%
\providecommand \@ifnum [1]{%
 \ifnum #1\expandafter \@firstoftwo
 \else \expandafter \@secondoftwo
 \fi
}%
\providecommand \@ifx [1]{%
 \ifx #1\expandafter \@firstoftwo
 \else \expandafter \@secondoftwo
 \fi
}%
\providecommand \natexlab [1]{#1}%
\providecommand \enquote  [1]{``#1''}%
\providecommand \bibnamefont  [1]{#1}%
\providecommand \bibfnamefont [1]{#1}%
\providecommand \citenamefont [1]{#1}%
\providecommand \href@noop [0]{\@secondoftwo}%
\providecommand \href [0]{\begingroup \@sanitize@url \@href}%
\providecommand \@href[1]{\@@startlink{#1}\@@href}%
\providecommand \@@href[1]{\endgroup#1\@@endlink}%
\providecommand \@sanitize@url [0]{\catcode `\\12\catcode `\$12\catcode
  `\&12\catcode `\#12\catcode `\^12\catcode `\_12\catcode `\%12\relax}%
\providecommand \@@startlink[1]{}%
\providecommand \@@endlink[0]{}%
\providecommand \url  [0]{\begingroup\@sanitize@url \@url }%
\providecommand \@url [1]{\endgroup\@href {#1}{\urlprefix }}%
\providecommand \urlprefix  [0]{URL }%
\providecommand \Eprint [0]{\href }%
\providecommand \doibase [0]{https://doi.org/}%
\providecommand \selectlanguage [0]{\@gobble}%
\providecommand \bibinfo  [0]{\@secondoftwo}%
\providecommand \bibfield  [0]{\@secondoftwo}%
\providecommand \translation [1]{[#1]}%
\providecommand \BibitemOpen [0]{}%
\providecommand \bibitemStop [0]{}%
\providecommand \bibitemNoStop [0]{.\EOS\space}%
\providecommand \EOS [0]{\spacefactor3000\relax}%
\providecommand \BibitemShut  [1]{\csname bibitem#1\endcsname}%
\let\auto@bib@innerbib\@empty
\bibitem [{\citenamefont {Jungwirth}\ \emph {et~al.}(2002)\citenamefont
  {Jungwirth}, \citenamefont {Niu},\ and\ \citenamefont
  {MacDonald}}]{FerroSemi02}%
  \BibitemOpen
  \bibfield  {author} {\bibinfo {author} {\bibfnamefont {T.}~\bibnamefont
  {Jungwirth}}, \bibinfo {author} {\bibfnamefont {Q.}~\bibnamefont {Niu}},\
  and\ \bibinfo {author} {\bibfnamefont {A.~H.}\ \bibnamefont {MacDonald}},\
  }\bibfield  {title} {\bibinfo {title} {Anomalous hall effect in ferromagnetic
  semiconductors},\ }\href {https://doi.org/10.1103/PhysRevLett.88.207208}
  {\bibfield  {journal} {\bibinfo  {journal} {Phys. Rev. Lett.}\ }\textbf
  {\bibinfo {volume} {88}},\ \bibinfo {pages} {207208} (\bibinfo {year}
  {2002})}\BibitemShut {NoStop}%
\bibitem [{\citenamefont {Nagaosa}\ \emph {et~al.}(2010)\citenamefont
  {Nagaosa}, \citenamefont {Sinova}, \citenamefont {Onoda}, \citenamefont
  {MacDonald},\ and\ \citenamefont {Ong}}]{AHE_RMP10}%
  \BibitemOpen
  \bibfield  {author} {\bibinfo {author} {\bibfnamefont {N.}~\bibnamefont
  {Nagaosa}}, \bibinfo {author} {\bibfnamefont {J.}~\bibnamefont {Sinova}},
  \bibinfo {author} {\bibfnamefont {S.}~\bibnamefont {Onoda}}, \bibinfo
  {author} {\bibfnamefont {A.~H.}\ \bibnamefont {MacDonald}},\ and\ \bibinfo
  {author} {\bibfnamefont {N.~P.}\ \bibnamefont {Ong}},\ }\bibfield  {title}
  {\bibinfo {title} {Anomalous hall effect},\ }\href
  {https://doi.org/10.1103/RevModPhys.82.1539} {\bibfield  {journal} {\bibinfo
  {journal} {Rev. Mod. Phys.}\ }\textbf {\bibinfo {volume} {82}},\ \bibinfo
  {pages} {1539} (\bibinfo {year} {2010})}\BibitemShut {NoStop}%
\bibitem [{\citenamefont {Weng}\ \emph {et~al.}(2015)\citenamefont {Weng},
  \citenamefont {Yu}, \citenamefont {Hu}, \citenamefont {Dai},\ and\
  \citenamefont {Fang}}]{Weng-2015}%
  \BibitemOpen
  \bibfield  {author} {\bibinfo {author} {\bibfnamefont {H.}~\bibnamefont
  {Weng}}, \bibinfo {author} {\bibfnamefont {R.}~\bibnamefont {Yu}}, \bibinfo
  {author} {\bibfnamefont {X.}~\bibnamefont {Hu}}, \bibinfo {author}
  {\bibfnamefont {X.}~\bibnamefont {Dai}},\ and\ \bibinfo {author}
  {\bibfnamefont {Z.}~\bibnamefont {Fang}},\ }\bibfield  {title} {\bibinfo
  {title} {Quantum anomalous hall effect and related topological electronic
  states},\ }\href {https://doi.org/10.1080/00018732.2015.1068524} {\bibfield
  {journal} {\bibinfo  {journal} {Advances in Physics}\ }\textbf {\bibinfo
  {volume} {64}},\ \bibinfo {pages} {227} (\bibinfo {year} {2015})},\ \Eprint
  {https://arxiv.org/abs/https://doi.org/10.1080/00018732.2015.1068524}
  {https://doi.org/10.1080/00018732.2015.1068524} \BibitemShut {NoStop}%
\bibitem [{\citenamefont {Liu}\ \emph {et~al.}(2016)\citenamefont {Liu},
  \citenamefont {Zhang},\ and\ \citenamefont {Qi}}]{Liu-2016}%
  \BibitemOpen
  \bibfield  {author} {\bibinfo {author} {\bibfnamefont {C.-X.}\ \bibnamefont
  {Liu}}, \bibinfo {author} {\bibfnamefont {S.-C.}\ \bibnamefont {Zhang}},\
  and\ \bibinfo {author} {\bibfnamefont {X.-L.}\ \bibnamefont {Qi}},\
  }\bibfield  {title} {\bibinfo {title} {The quantum anomalous hall effect:
  Theory and experiment},\ }\href
  {https://doi.org/10.1146/annurev-conmatphys-031115-011417} {\bibfield
  {journal} {\bibinfo  {journal} {Annual Review of Condensed Matter Physics}\
  }\textbf {\bibinfo {volume} {7}},\ \bibinfo {pages} {301} (\bibinfo {year}
  {2016})},\ \Eprint
  {https://arxiv.org/abs/https://doi.org/10.1146/annurev-conmatphys-031115-011417}
  {https://doi.org/10.1146/annurev-conmatphys-031115-011417} \BibitemShut
  {NoStop}%
\bibitem [{\citenamefont {Chang}\ and\ \citenamefont {Li}(2016)}]{Chang-2016}%
  \BibitemOpen
  \bibfield  {author} {\bibinfo {author} {\bibfnamefont {C.-Z.}\ \bibnamefont
  {Chang}}\ and\ \bibinfo {author} {\bibfnamefont {M.}~\bibnamefont {Li}},\
  }\href {https://doi.org/doi:10.1088/0953-8984/28/12/123002} {\bibfield
  {journal} {\bibinfo  {journal} {J. Phys.: Condens. Matter}\ }\textbf
  {\bibinfo {volume} {28}},\ \bibinfo {pages} {123002} (\bibinfo {year}
  {2016})}\BibitemShut {NoStop}%
\bibitem [{\citenamefont {He}\ \emph {et~al.}(2018)\citenamefont {He},
  \citenamefont {Wang},\ and\ \citenamefont {Xue}}]{He-2018}%
  \BibitemOpen
  \bibfield  {author} {\bibinfo {author} {\bibfnamefont {K.}~\bibnamefont
  {He}}, \bibinfo {author} {\bibfnamefont {Y.}~\bibnamefont {Wang}},\ and\
  \bibinfo {author} {\bibfnamefont {Q.-K.}\ \bibnamefont {Xue}},\ }\bibfield
  {title} {\bibinfo {title} {Topological materials: Quantum anomalous hall
  system},\ }\href {https://doi.org/10.1146/annurev-conmatphys-033117-054144}
  {\bibfield  {journal} {\bibinfo  {journal} {Annual Review of Condensed Matter
  Physics}\ }\textbf {\bibinfo {volume} {9}},\ \bibinfo {pages} {329} (\bibinfo
  {year} {2018})},\ \Eprint
  {https://arxiv.org/abs/https://doi.org/10.1146/annurev-conmatphys-033117-054144}
  {https://doi.org/10.1146/annurev-conmatphys-033117-054144} \BibitemShut
  {NoStop}%
\bibitem [{\citenamefont {Tokura}\ \emph {et~al.}(2019)\citenamefont {Tokura},
  \citenamefont {Yasuda},\ and\ \citenamefont {Tsukazaki}}]{NRP-2019}%
  \BibitemOpen
  \bibfield  {author} {\bibinfo {author} {\bibfnamefont {Y.}~\bibnamefont
  {Tokura}}, \bibinfo {author} {\bibfnamefont {K.}~\bibnamefont {Yasuda}},\
  and\ \bibinfo {author} {\bibfnamefont {A.}~\bibnamefont {Tsukazaki}},\
  }\bibfield  {title} {\bibinfo {title} {Magnetic topological insulators},\
  }\href {https://doi.org/10.1038/s42254-018-0011-5} {\bibfield  {journal}
  {\bibinfo  {journal} {Nature Reviews Physics}\ }\textbf {\bibinfo {volume}
  {1}},\ \bibinfo {pages} {126} (\bibinfo {year} {2019})}\BibitemShut {NoStop}%
\bibitem [{\citenamefont {Chang}\ \emph {et~al.}(2022)\citenamefont {Chang},
  \citenamefont {Liu},\ and\ \citenamefont {MacDonald}}]{Chang-2022}%
  \BibitemOpen
  \bibfield  {author} {\bibinfo {author} {\bibfnamefont {C.-Z.}\ \bibnamefont
  {Chang}}, \bibinfo {author} {\bibfnamefont {C.-X.}\ \bibnamefont {Liu}},\
  and\ \bibinfo {author} {\bibfnamefont {A.~H.}\ \bibnamefont {MacDonald}},\
  }\href {https://doi.org/arxiv.org/abs/2202.13902} {\bibfield  {journal}
  {\bibinfo  {journal} {ArXiv:}\ }\textbf {\bibinfo {volume} {2202}},\ \bibinfo
  {pages} {13902} (\bibinfo {year} {2022})}\BibitemShut {NoStop}%
\bibitem [{\citenamefont {Haldane}(1988)}]{Haldane88}%
  \BibitemOpen
  \bibfield  {author} {\bibinfo {author} {\bibfnamefont {F.~D.~M.}\
  \bibnamefont {Haldane}},\ }\bibfield  {title} {\bibinfo {title} {Model for a
  quantum hall effect without landau levels: Condensed-matter realization of
  the "parity anomaly"},\ }\href {https://doi.org/10.1103/PhysRevLett.61.2015}
  {\bibfield  {journal} {\bibinfo  {journal} {Phys. Rev. Lett.}\ }\textbf
  {\bibinfo {volume} {61}},\ \bibinfo {pages} {2015} (\bibinfo {year}
  {1988})}\BibitemShut {NoStop}%
\bibitem [{\citenamefont {Qi}\ \emph {et~al.}(2006)\citenamefont {Qi},
  \citenamefont {Wu},\ and\ \citenamefont {Zhang}}]{QiWuZhang06}%
  \BibitemOpen
  \bibfield  {author} {\bibinfo {author} {\bibfnamefont {X.-L.}\ \bibnamefont
  {Qi}}, \bibinfo {author} {\bibfnamefont {Y.-S.}\ \bibnamefont {Wu}},\ and\
  \bibinfo {author} {\bibfnamefont {S.-C.}\ \bibnamefont {Zhang}},\ }\bibfield
  {title} {\bibinfo {title} {Topological quantization of the spin hall effect
  in two-dimensional paramagnetic semiconductors},\ }\href
  {https://doi.org/10.1103/PhysRevB.74.085308} {\bibfield  {journal} {\bibinfo
  {journal} {Phys. Rev. B}\ }\textbf {\bibinfo {volume} {74}},\ \bibinfo
  {pages} {085308} (\bibinfo {year} {2006})}\BibitemShut {NoStop}%
\bibitem [{\citenamefont {Jiang}\ \emph {et~al.}(2018)\citenamefont {Jiang},
  \citenamefont {Zhou}, \citenamefont {Dai},\ and\ \citenamefont
  {Wang}}]{Jiang-2018}%
  \BibitemOpen
  \bibfield  {author} {\bibinfo {author} {\bibfnamefont {K.}~\bibnamefont
  {Jiang}}, \bibinfo {author} {\bibfnamefont {S.}~\bibnamefont {Zhou}},
  \bibinfo {author} {\bibfnamefont {X.}~\bibnamefont {Dai}},\ and\ \bibinfo
  {author} {\bibfnamefont {Z.}~\bibnamefont {Wang}},\ }\bibfield  {title}
  {\bibinfo {title} {Antiferromagnetic chern insulators in noncentrosymmetric
  systems},\ }\href {https://doi.org/10.1103/PhysRevLett.120.157205} {\bibfield
   {journal} {\bibinfo  {journal} {Phys. Rev. Lett.}\ }\textbf {\bibinfo
  {volume} {120}},\ \bibinfo {pages} {157205} (\bibinfo {year}
  {2018})}\BibitemShut {NoStop}%
\bibitem [{\citenamefont {Liu}\ \emph {et~al.}(2008)\citenamefont {Liu},
  \citenamefont {Qi}, \citenamefont {Dai}, \citenamefont {Fang},\ and\
  \citenamefont {Zhang}}]{Liu-2008}%
  \BibitemOpen
  \bibfield  {author} {\bibinfo {author} {\bibfnamefont {C.-X.}\ \bibnamefont
  {Liu}}, \bibinfo {author} {\bibfnamefont {X.-L.}\ \bibnamefont {Qi}},
  \bibinfo {author} {\bibfnamefont {X.}~\bibnamefont {Dai}}, \bibinfo {author}
  {\bibfnamefont {Z.}~\bibnamefont {Fang}},\ and\ \bibinfo {author}
  {\bibfnamefont {S.-C.}\ \bibnamefont {Zhang}},\ }\bibfield  {title} {\bibinfo
  {title} {Quantum anomalous hall effect in
  ${\mathrm{hg}}_{1\ensuremath{-}y}{\mathrm{mn}}_{y}\mathrm{Te}$ quantum
  wells},\ }\href {https://doi.org/10.1103/PhysRevLett.101.146802} {\bibfield
  {journal} {\bibinfo  {journal} {Phys. Rev. Lett.}\ }\textbf {\bibinfo
  {volume} {101}},\ \bibinfo {pages} {146802} (\bibinfo {year}
  {2008})}\BibitemShut {NoStop}%
\bibitem [{\citenamefont {Yu}\ \emph {et~al.}(2010)\citenamefont {Yu},
  \citenamefont {Zhang}, \citenamefont {Zhang}, \citenamefont {Zhang},
  \citenamefont {Dai},\ and\ \citenamefont {Fang}}]{yu-2010}%
  \BibitemOpen
  \bibfield  {author} {\bibinfo {author} {\bibfnamefont {R.}~\bibnamefont
  {Yu}}, \bibinfo {author} {\bibfnamefont {W.}~\bibnamefont {Zhang}}, \bibinfo
  {author} {\bibfnamefont {H.-J.}\ \bibnamefont {Zhang}}, \bibinfo {author}
  {\bibfnamefont {S.-C.}\ \bibnamefont {Zhang}}, \bibinfo {author}
  {\bibfnamefont {X.}~\bibnamefont {Dai}},\ and\ \bibinfo {author}
  {\bibfnamefont {Z.}~\bibnamefont {Fang}},\ }\bibfield  {title} {\bibinfo
  {title} {Quantized anomalous hall effect in magnetic topological
  insulators},\ }\href {https://doi.org/10.1126/science.1187485} {\bibfield
  {journal} {\bibinfo  {journal} {Science}\ }\textbf {\bibinfo {volume}
  {329}},\ \bibinfo {pages} {61} (\bibinfo {year} {2010})}\BibitemShut
  {NoStop}%
\bibitem [{\citenamefont {Wang}\ \emph {et~al.}(2013)\citenamefont {Wang},
  \citenamefont {Liu},\ and\ \citenamefont {Liu}}]{Wang-2013}%
  \BibitemOpen
  \bibfield  {author} {\bibinfo {author} {\bibfnamefont {Z.~F.}\ \bibnamefont
  {Wang}}, \bibinfo {author} {\bibfnamefont {Z.}~\bibnamefont {Liu}},\ and\
  \bibinfo {author} {\bibfnamefont {F.}~\bibnamefont {Liu}},\ }\bibfield
  {title} {\bibinfo {title} {Quantum anomalous hall effect in 2d organic
  topological insulators},\ }\href
  {https://doi.org/10.1103/PhysRevLett.110.196801} {\bibfield  {journal}
  {\bibinfo  {journal} {Phys. Rev. Lett.}\ }\textbf {\bibinfo {volume} {110}},\
  \bibinfo {pages} {196801} (\bibinfo {year} {2013})}\BibitemShut {NoStop}%
\bibitem [{\citenamefont {Fang}\ \emph {et~al.}(2014)\citenamefont {Fang},
  \citenamefont {Gilbert},\ and\ \citenamefont {Bernevig}}]{Fang-2014}%
  \BibitemOpen
  \bibfield  {author} {\bibinfo {author} {\bibfnamefont {C.}~\bibnamefont
  {Fang}}, \bibinfo {author} {\bibfnamefont {M.~J.}\ \bibnamefont {Gilbert}},\
  and\ \bibinfo {author} {\bibfnamefont {B.~A.}\ \bibnamefont {Bernevig}},\
  }\bibfield  {title} {\bibinfo {title} {Large-chern-number quantum anomalous
  hall effect in thin-film topological crystalline insulators},\ }\href
  {https://doi.org/10.1103/PhysRevLett.112.046801} {\bibfield  {journal}
  {\bibinfo  {journal} {Phys. Rev. Lett.}\ }\textbf {\bibinfo {volume} {112}},\
  \bibinfo {pages} {046801} (\bibinfo {year} {2014})}\BibitemShut {NoStop}%
\bibitem [{\citenamefont {Wu}\ \emph {et~al.}(2014)\citenamefont {Wu},
  \citenamefont {Shan},\ and\ \citenamefont {Yan}}]{Wu-2014}%
  \BibitemOpen
  \bibfield  {author} {\bibinfo {author} {\bibfnamefont {S.-C.}\ \bibnamefont
  {Wu}}, \bibinfo {author} {\bibfnamefont {G.}~\bibnamefont {Shan}},\ and\
  \bibinfo {author} {\bibfnamefont {B.}~\bibnamefont {Yan}},\ }\bibfield
  {title} {\bibinfo {title} {Prediction of near-room-temperature quantum
  anomalous hall effect on honeycomb materials},\ }\href
  {https://doi.org/10.1103/PhysRevLett.113.256401} {\bibfield  {journal}
  {\bibinfo  {journal} {Phys. Rev. Lett.}\ }\textbf {\bibinfo {volume} {113}},\
  \bibinfo {pages} {256401} (\bibinfo {year} {2014})}\BibitemShut {NoStop}%
\bibitem [{\citenamefont {Ren}\ \emph {et~al.}(2016)\citenamefont {Ren},
  \citenamefont {Zeng}, \citenamefont {Deng}, \citenamefont {Yang},
  \citenamefont {Pan},\ and\ \citenamefont {Qiao}}]{Ren-2016}%
  \BibitemOpen
  \bibfield  {author} {\bibinfo {author} {\bibfnamefont {Y.}~\bibnamefont
  {Ren}}, \bibinfo {author} {\bibfnamefont {J.}~\bibnamefont {Zeng}}, \bibinfo
  {author} {\bibfnamefont {X.}~\bibnamefont {Deng}}, \bibinfo {author}
  {\bibfnamefont {F.}~\bibnamefont {Yang}}, \bibinfo {author} {\bibfnamefont
  {H.}~\bibnamefont {Pan}},\ and\ \bibinfo {author} {\bibfnamefont
  {Z.}~\bibnamefont {Qiao}},\ }\bibfield  {title} {\bibinfo {title} {Quantum
  anomalous hall effect in atomic crystal layers from in-plane magnetization},\
  }\href {https://doi.org/10.1103/PhysRevB.94.085411} {\bibfield  {journal}
  {\bibinfo  {journal} {Phys. Rev. B}\ }\textbf {\bibinfo {volume} {94}},\
  \bibinfo {pages} {085411} (\bibinfo {year} {2016})}\BibitemShut {NoStop}%
\bibitem [{\citenamefont {Huang}\ \emph {et~al.}(2017)\citenamefont {Huang},
  \citenamefont {Zhou}, \citenamefont {Wu}, \citenamefont {Deng}, \citenamefont
  {Jena},\ and\ \citenamefont {Kan}}]{Huang-2017}%
  \BibitemOpen
  \bibfield  {author} {\bibinfo {author} {\bibfnamefont {C.}~\bibnamefont
  {Huang}}, \bibinfo {author} {\bibfnamefont {J.}~\bibnamefont {Zhou}},
  \bibinfo {author} {\bibfnamefont {H.}~\bibnamefont {Wu}}, \bibinfo {author}
  {\bibfnamefont {K.}~\bibnamefont {Deng}}, \bibinfo {author} {\bibfnamefont
  {P.}~\bibnamefont {Jena}},\ and\ \bibinfo {author} {\bibfnamefont
  {E.}~\bibnamefont {Kan}},\ }\bibfield  {title} {\bibinfo {title} {Quantum
  anomalous hall effect in ferromagnetic transition metal halides},\ }\href
  {https://doi.org/10.1103/PhysRevB.95.045113} {\bibfield  {journal} {\bibinfo
  {journal} {Phys. Rev. B}\ }\textbf {\bibinfo {volume} {95}},\ \bibinfo
  {pages} {045113} (\bibinfo {year} {2017})}\BibitemShut {NoStop}%
\bibitem [{\citenamefont {Chen}\ \emph {et~al.}(2017)\citenamefont {Chen},
  \citenamefont {Zou},\ and\ \citenamefont {Liu}}]{Chen-2017}%
  \BibitemOpen
  \bibfield  {author} {\bibinfo {author} {\bibfnamefont {P.}~\bibnamefont
  {Chen}}, \bibinfo {author} {\bibfnamefont {J.-Y.}\ \bibnamefont {Zou}},\ and\
  \bibinfo {author} {\bibfnamefont {B.-G.}\ \bibnamefont {Liu}},\ }\bibfield
  {title} {\bibinfo {title} {Intrinsic ferromagnetism and quantum anomalous
  hall effect in a cobr2 monolayer},\ }\href
  {https://doi.org/10.1039/C7CP02158E} {\bibfield  {journal} {\bibinfo
  {journal} {Phys. Chem. Chem. Phys.}\ }\textbf {\bibinfo {volume} {19}},\
  \bibinfo {pages} {13432} (\bibinfo {year} {2017})}\BibitemShut {NoStop}%
\bibitem [{\citenamefont {You}\ \emph {et~al.}(2019)\citenamefont {You},
  \citenamefont {Zhang}, \citenamefont {Gu},\ and\ \citenamefont
  {Su}}]{You-2019}%
  \BibitemOpen
  \bibfield  {author} {\bibinfo {author} {\bibfnamefont {J.-Y.}\ \bibnamefont
  {You}}, \bibinfo {author} {\bibfnamefont {Z.}~\bibnamefont {Zhang}}, \bibinfo
  {author} {\bibfnamefont {B.}~\bibnamefont {Gu}},\ and\ \bibinfo {author}
  {\bibfnamefont {G.}~\bibnamefont {Su}},\ }\bibfield  {title} {\bibinfo
  {title} {Two-dimensional room-temperature ferromagnetic semiconductors with
  quantum anomalous hall effect},\ }\href
  {https://doi.org/10.1103/PhysRevApplied.12.024063} {\bibfield  {journal}
  {\bibinfo  {journal} {Phys. Rev. Applied}\ }\textbf {\bibinfo {volume}
  {12}},\ \bibinfo {pages} {024063} (\bibinfo {year} {2019})}\BibitemShut
  {NoStop}%
\bibitem [{\citenamefont {Wu}\ \emph {et~al.}(2019)\citenamefont {Wu},
  \citenamefont {Lovorn}, \citenamefont {Tutuc}, \citenamefont {Martin},\ and\
  \citenamefont {MacDonald}}]{Wu-2019}%
  \BibitemOpen
  \bibfield  {author} {\bibinfo {author} {\bibfnamefont {F.}~\bibnamefont
  {Wu}}, \bibinfo {author} {\bibfnamefont {T.}~\bibnamefont {Lovorn}}, \bibinfo
  {author} {\bibfnamefont {E.}~\bibnamefont {Tutuc}}, \bibinfo {author}
  {\bibfnamefont {I.}~\bibnamefont {Martin}},\ and\ \bibinfo {author}
  {\bibfnamefont {A.~H.}\ \bibnamefont {MacDonald}},\ }\bibfield  {title}
  {\bibinfo {title} {Topological insulators in twisted transition metal
  dichalcogenide homobilayers},\ }\href
  {https://doi.org/10.1103/PhysRevLett.122.086402} {\bibfield  {journal}
  {\bibinfo  {journal} {Phys. Rev. Lett.}\ }\textbf {\bibinfo {volume} {122}},\
  \bibinfo {pages} {086402} (\bibinfo {year} {2019})}\BibitemShut {NoStop}%
\bibitem [{\citenamefont {Zhang}\ \emph
  {et~al.}(2019{\natexlab{a}})\citenamefont {Zhang}, \citenamefont {Shi},
  \citenamefont {Zhu}, \citenamefont {Xing}, \citenamefont {Zhang},\ and\
  \citenamefont {Wang}}]{Zhang-2019}%
  \BibitemOpen
  \bibfield  {author} {\bibinfo {author} {\bibfnamefont {D.}~\bibnamefont
  {Zhang}}, \bibinfo {author} {\bibfnamefont {M.}~\bibnamefont {Shi}}, \bibinfo
  {author} {\bibfnamefont {T.}~\bibnamefont {Zhu}}, \bibinfo {author}
  {\bibfnamefont {D.}~\bibnamefont {Xing}}, \bibinfo {author} {\bibfnamefont
  {H.}~\bibnamefont {Zhang}},\ and\ \bibinfo {author} {\bibfnamefont
  {J.}~\bibnamefont {Wang}},\ }\bibfield  {title} {\bibinfo {title}
  {Topological axion states in the magnetic insulator
  ${\mathrm{mnbi}}_{2}{\mathrm{te}}_{4}$ with the quantized magnetoelectric
  effect},\ }\href {https://doi.org/10.1103/PhysRevLett.122.206401} {\bibfield
  {journal} {\bibinfo  {journal} {Phys. Rev. Lett.}\ }\textbf {\bibinfo
  {volume} {122}},\ \bibinfo {pages} {206401} (\bibinfo {year}
  {2019}{\natexlab{a}})}\BibitemShut {NoStop}%
\bibitem [{\citenamefont {Zhang}\ \emph
  {et~al.}(2019{\natexlab{b}})\citenamefont {Zhang}, \citenamefont {Mao},\ and\
  \citenamefont {Senthil}}]{PRR-2019}%
  \BibitemOpen
  \bibfield  {author} {\bibinfo {author} {\bibfnamefont {Y.-H.}\ \bibnamefont
  {Zhang}}, \bibinfo {author} {\bibfnamefont {D.}~\bibnamefont {Mao}},\ and\
  \bibinfo {author} {\bibfnamefont {T.}~\bibnamefont {Senthil}},\ }\bibfield
  {title} {\bibinfo {title} {Twisted bilayer graphene aligned with hexagonal
  boron nitride: Anomalous hall effect and a lattice model},\ }\href
  {https://doi.org/10.1103/PhysRevResearch.1.033126} {\bibfield  {journal}
  {\bibinfo  {journal} {Phys. Rev. Research}\ }\textbf {\bibinfo {volume}
  {1}},\ \bibinfo {pages} {033126} (\bibinfo {year}
  {2019}{\natexlab{b}})}\BibitemShut {NoStop}%
\bibitem [{\citenamefont {Zhang}\ \emph
  {et~al.}(2019{\natexlab{c}})\citenamefont {Zhang}, \citenamefont {Mao},
  \citenamefont {Cao}, \citenamefont {Jarillo-Herrero},\ and\ \citenamefont
  {Senthil}}]{PRB-2019}%
  \BibitemOpen
  \bibfield  {author} {\bibinfo {author} {\bibfnamefont {Y.-H.}\ \bibnamefont
  {Zhang}}, \bibinfo {author} {\bibfnamefont {D.}~\bibnamefont {Mao}}, \bibinfo
  {author} {\bibfnamefont {Y.}~\bibnamefont {Cao}}, \bibinfo {author}
  {\bibfnamefont {P.}~\bibnamefont {Jarillo-Herrero}},\ and\ \bibinfo {author}
  {\bibfnamefont {T.}~\bibnamefont {Senthil}},\ }\bibfield  {title} {\bibinfo
  {title} {Nearly flat chern bands in moir\'e superlattices},\ }\href
  {https://doi.org/10.1103/PhysRevB.99.075127} {\bibfield  {journal} {\bibinfo
  {journal} {Phys. Rev. B}\ }\textbf {\bibinfo {volume} {99}},\ \bibinfo
  {pages} {075127} (\bibinfo {year} {2019}{\natexlab{c}})}\BibitemShut
  {NoStop}%
\bibitem [{\citenamefont {Bultinck}\ \emph {et~al.}(2020)\citenamefont
  {Bultinck}, \citenamefont {Chatterjee},\ and\ \citenamefont
  {Zaletel}}]{PRL-2020}%
  \BibitemOpen
  \bibfield  {author} {\bibinfo {author} {\bibfnamefont {N.}~\bibnamefont
  {Bultinck}}, \bibinfo {author} {\bibfnamefont {S.}~\bibnamefont
  {Chatterjee}},\ and\ \bibinfo {author} {\bibfnamefont {M.~P.}\ \bibnamefont
  {Zaletel}},\ }\bibfield  {title} {\bibinfo {title} {Mechanism for anomalous
  hall ferromagnetism in twisted bilayer graphene},\ }\href
  {https://doi.org/10.1103/PhysRevLett.124.166601} {\bibfield  {journal}
  {\bibinfo  {journal} {Phys. Rev. Lett.}\ }\textbf {\bibinfo {volume} {124}},\
  \bibinfo {pages} {166601} (\bibinfo {year} {2020})}\BibitemShut {NoStop}%
\bibitem [{\citenamefont {Shi}\ \emph {et~al.}(2021)\citenamefont {Shi},
  \citenamefont {Zhu},\ and\ \citenamefont {MacDonald}}]{Shi-2021}%
  \BibitemOpen
  \bibfield  {author} {\bibinfo {author} {\bibfnamefont {J.}~\bibnamefont
  {Shi}}, \bibinfo {author} {\bibfnamefont {J.}~\bibnamefont {Zhu}},\ and\
  \bibinfo {author} {\bibfnamefont {A.~H.}\ \bibnamefont {MacDonald}},\
  }\bibfield  {title} {\bibinfo {title} {Moir\'e commensurability and the
  quantum anomalous hall effect in twisted bilayer graphene on hexagonal boron
  nitride},\ }\href {https://doi.org/10.1103/PhysRevB.103.075122} {\bibfield
  {journal} {\bibinfo  {journal} {Phys. Rev. B}\ }\textbf {\bibinfo {volume}
  {103}},\ \bibinfo {pages} {075122} (\bibinfo {year} {2021})}\BibitemShut
  {NoStop}%
\bibitem [{\citenamefont {Chang}\ \emph {et~al.}(2013)\citenamefont {Chang},
  \citenamefont {Zhang}, \citenamefont {Feng}, \citenamefont {Shen},
  \citenamefont {Zhang}, \citenamefont {Guo}, \citenamefont {Li}, \citenamefont
  {Ou}, \citenamefont {Wei}, \citenamefont {Wang}, \citenamefont {Ji},
  \citenamefont {Feng}, \citenamefont {Ji}, \citenamefont {Chen}, \citenamefont
  {Jia}, \citenamefont {Dai}, \citenamefont {Fang}, \citenamefont {Zhang},
  \citenamefont {He}, \citenamefont {Wang}, \citenamefont {Lu}, \citenamefont
  {Ma},\ and\ \citenamefont {Xue}}]{chang-2013}%
  \BibitemOpen
  \bibfield  {author} {\bibinfo {author} {\bibfnamefont {C.-Z.}\ \bibnamefont
  {Chang}}, \bibinfo {author} {\bibfnamefont {J.}~\bibnamefont {Zhang}},
  \bibinfo {author} {\bibfnamefont {X.}~\bibnamefont {Feng}}, \bibinfo {author}
  {\bibfnamefont {J.}~\bibnamefont {Shen}}, \bibinfo {author} {\bibfnamefont
  {Z.}~\bibnamefont {Zhang}}, \bibinfo {author} {\bibfnamefont
  {M.}~\bibnamefont {Guo}}, \bibinfo {author} {\bibfnamefont {K.}~\bibnamefont
  {Li}}, \bibinfo {author} {\bibfnamefont {Y.}~\bibnamefont {Ou}}, \bibinfo
  {author} {\bibfnamefont {P.}~\bibnamefont {Wei}}, \bibinfo {author}
  {\bibfnamefont {L.-L.}\ \bibnamefont {Wang}}, \bibinfo {author}
  {\bibfnamefont {Z.-Q.}\ \bibnamefont {Ji}}, \bibinfo {author} {\bibfnamefont
  {Y.}~\bibnamefont {Feng}}, \bibinfo {author} {\bibfnamefont {S.}~\bibnamefont
  {Ji}}, \bibinfo {author} {\bibfnamefont {X.}~\bibnamefont {Chen}}, \bibinfo
  {author} {\bibfnamefont {J.}~\bibnamefont {Jia}}, \bibinfo {author}
  {\bibfnamefont {X.}~\bibnamefont {Dai}}, \bibinfo {author} {\bibfnamefont
  {Z.}~\bibnamefont {Fang}}, \bibinfo {author} {\bibfnamefont {S.-C.}\
  \bibnamefont {Zhang}}, \bibinfo {author} {\bibfnamefont {K.}~\bibnamefont
  {He}}, \bibinfo {author} {\bibfnamefont {Y.}~\bibnamefont {Wang}}, \bibinfo
  {author} {\bibfnamefont {L.}~\bibnamefont {Lu}}, \bibinfo {author}
  {\bibfnamefont {X.-C.}\ \bibnamefont {Ma}},\ and\ \bibinfo {author}
  {\bibfnamefont {Q.-K.}\ \bibnamefont {Xue}},\ }\bibfield  {title} {\bibinfo
  {title} {Experimental observation of the quantum anomalous hall effect in a
  magnetic topological insulator},\ }\href
  {https://doi.org/10.1126/science.1234414} {\bibfield  {journal} {\bibinfo
  {journal} {Science}\ }\textbf {\bibinfo {volume} {340}},\ \bibinfo {pages}
  {167} (\bibinfo {year} {2013})}\BibitemShut {NoStop}%
\bibitem [{\citenamefont {Deng}\ \emph {et~al.}(2020)\citenamefont {Deng},
  \citenamefont {Yu}, \citenamefont {Shi}, \citenamefont {Guo}, \citenamefont
  {Xu}, \citenamefont {Wang}, \citenamefont {Chen},\ and\ \citenamefont
  {Zhang}}]{Deng-2020}%
  \BibitemOpen
  \bibfield  {author} {\bibinfo {author} {\bibfnamefont {Y.}~\bibnamefont
  {Deng}}, \bibinfo {author} {\bibfnamefont {Y.}~\bibnamefont {Yu}}, \bibinfo
  {author} {\bibfnamefont {M.~Z.}\ \bibnamefont {Shi}}, \bibinfo {author}
  {\bibfnamefont {Z.}~\bibnamefont {Guo}}, \bibinfo {author} {\bibfnamefont
  {Z.}~\bibnamefont {Xu}}, \bibinfo {author} {\bibfnamefont {J.}~\bibnamefont
  {Wang}}, \bibinfo {author} {\bibfnamefont {X.~H.}\ \bibnamefont {Chen}},\
  and\ \bibinfo {author} {\bibfnamefont {Y.}~\bibnamefont {Zhang}},\ }\bibfield
   {title} {\bibinfo {title} {Quantum anomalous hall effect in intrinsic
  magnetic topological insulator mnbi(2)te(4)},\ }\href
  {https://doi.org/10.1126/science.aax8156} {\bibfield  {journal} {\bibinfo
  {journal} {Science}\ }\textbf {\bibinfo {volume} {367}},\ \bibinfo {pages}
  {895} (\bibinfo {year} {2020})}\BibitemShut {NoStop}%
\bibitem [{\citenamefont {Serlin}\ \emph {et~al.}(2020)\citenamefont {Serlin},
  \citenamefont {Tschirhart}, \citenamefont {Polshyn}, \citenamefont {Zhang},
  \citenamefont {Zhu}, \citenamefont {Watanabe}, \citenamefont {Taniguchi},
  \citenamefont {Balents},\ and\ \citenamefont {Young}}]{Science-2020}%
  \BibitemOpen
  \bibfield  {author} {\bibinfo {author} {\bibfnamefont {M.}~\bibnamefont
  {Serlin}}, \bibinfo {author} {\bibfnamefont {C.~L.}\ \bibnamefont
  {Tschirhart}}, \bibinfo {author} {\bibfnamefont {H.}~\bibnamefont {Polshyn}},
  \bibinfo {author} {\bibfnamefont {Y.}~\bibnamefont {Zhang}}, \bibinfo
  {author} {\bibfnamefont {J.}~\bibnamefont {Zhu}}, \bibinfo {author}
  {\bibfnamefont {K.}~\bibnamefont {Watanabe}}, \bibinfo {author}
  {\bibfnamefont {T.}~\bibnamefont {Taniguchi}}, \bibinfo {author}
  {\bibfnamefont {L.}~\bibnamefont {Balents}},\ and\ \bibinfo {author}
  {\bibfnamefont {A.~F.~a.}\ \bibnamefont {Young}},\ }\bibfield  {title}
  {\bibinfo {title} {Intrinsic quantized anomalous hall effect in a moir\'{e}
  heterostructure},\ }\href {https://doi.org/doi: 10.1126/science.aay5533}
  {\bibfield  {journal} {\bibinfo  {journal} {Science}\ }\textbf {\bibinfo
  {volume} {367}},\ \bibinfo {pages} {900} (\bibinfo {year}
  {2020})}\BibitemShut {NoStop}%
\bibitem [{\citenamefont {Li}\ \emph {et~al.}(2021)\citenamefont {Li},
  \citenamefont {Jiang}, \citenamefont {Shen}, \citenamefont {Zhang},
  \citenamefont {Li}, \citenamefont {Tao}, \citenamefont {Devakul},
  \citenamefont {Watanabe}, \citenamefont {Taniguchi}, \citenamefont {Fu},
  \citenamefont {Shan},\ and\ \citenamefont {Mak}}]{Nature-2020}%
  \BibitemOpen
  \bibfield  {author} {\bibinfo {author} {\bibfnamefont {T.}~\bibnamefont
  {Li}}, \bibinfo {author} {\bibfnamefont {S.}~\bibnamefont {Jiang}}, \bibinfo
  {author} {\bibfnamefont {B.}~\bibnamefont {Shen}}, \bibinfo {author}
  {\bibfnamefont {Y.}~\bibnamefont {Zhang}}, \bibinfo {author} {\bibfnamefont
  {L.}~\bibnamefont {Li}}, \bibinfo {author} {\bibfnamefont {Z.}~\bibnamefont
  {Tao}}, \bibinfo {author} {\bibfnamefont {T.}~\bibnamefont {Devakul}},
  \bibinfo {author} {\bibfnamefont {K.}~\bibnamefont {Watanabe}}, \bibinfo
  {author} {\bibfnamefont {T.}~\bibnamefont {Taniguchi}}, \bibinfo {author}
  {\bibfnamefont {L.}~\bibnamefont {Fu}}, \bibinfo {author} {\bibfnamefont
  {J.}~\bibnamefont {Shan}},\ and\ \bibinfo {author} {\bibfnamefont {K.~F.}\
  \bibnamefont {Mak}},\ }\bibfield  {title} {\bibinfo {title} {Quantum
  anomalous hall effect from intertwined moir\'{e} bands},\ }\href
  {https://doi.org/10.1038/s41586-021-04171-1} {\bibfield  {journal} {\bibinfo
  {journal} {Nature}\ }\textbf {\bibinfo {volume} {600}},\ \bibinfo {pages}
  {641} (\bibinfo {year} {2021})}\BibitemShut {NoStop}%
\bibitem [{\citenamefont {Litvin}\ and\ \citenamefont
  {Opechowski}(1974)}]{Litvin1974}%
  \BibitemOpen
  \bibfield  {author} {\bibinfo {author} {\bibfnamefont {D.}~\bibnamefont
  {Litvin}}\ and\ \bibinfo {author} {\bibfnamefont {W.}~\bibnamefont
  {Opechowski}},\ }\bibfield  {title} {
  {Spin groups},\ }\href{https://doi.org/10.1016/0031-8914(74)90157-8} {\bibfield  {journal}
  {\bibinfo  {journal} {Physica}\ }\textbf {\bibinfo {volume} {76}},\ \bibinfo
  {pages} {538} (\bibinfo {year} {1974})}.\BibitemShut {Stop}%
  \bibitem [{\citenamefont {Liu}\ \emph {et~al.}(2022)\citenamefont {Liu},
  \citenamefont {Li}, \citenamefont {Han}, \citenamefont {Wan},\ and\
  \citenamefont {Liu}}]{QHLiuXGWanprx}%
  \BibitemOpen
  \bibfield  {author} {\bibinfo {author} {\bibfnamefont {P.}~\bibnamefont
  {Liu}}, \bibinfo {author} {\bibfnamefont {J.}~\bibnamefont {Li}}, \bibinfo
  {author} {\bibfnamefont {J.}~\bibnamefont {Han}}, \bibinfo {author}
  {\bibfnamefont {X.}~\bibnamefont {Wan}},\ and\ \bibinfo {author}
  {\bibfnamefont {Q.}~\bibnamefont {Liu}},\ }\bibfield  {title} {\bibinfo
  {title} {Spin-group symmetry in magnetic materials with negligible spin-orbit
  coupling},\ }\href {https://doi.org/10.1103/PhysRevX.12.021016} {\bibfield
  {journal} {\bibinfo  {journal} {Phys. Rev. X}\ }\textbf {\bibinfo {volume}
  {12}},\ \bibinfo {pages} {021016} (\bibinfo {year} {2022})}\BibitemShut
  {NoStop}%
\bibitem [{\citenamefont {Guo}\ \emph {et~al.}(2021)\citenamefont {Guo},
  \citenamefont {Wei}, \citenamefont {Liu}, \citenamefont {Liu},\ and\
  \citenamefont {Lu}}]{GWLLL21}%
  \BibitemOpen
  \bibfield  {author} {\bibinfo {author} {\bibfnamefont {P.-J.}\ \bibnamefont
  {Guo}}, \bibinfo {author} {\bibfnamefont {Y.-W.}\ \bibnamefont {Wei}},
  \bibinfo {author} {\bibfnamefont {K.}~\bibnamefont {Liu}}, \bibinfo {author}
  {\bibfnamefont {Z.-X.}\ \bibnamefont {Liu}},\ and\ \bibinfo {author}
  {\bibfnamefont {Z.-Y.}\ \bibnamefont {Lu}},\ }\bibfield  {title} {\bibinfo
  {title} {Eightfold degenerate fermions in two dimensions},\ }\href
  {https://doi.org/10.1103/PhysRevLett.127.176401} {\bibfield  {journal}
  {\bibinfo  {journal} {Phys. Rev. Lett.}\ }\textbf {\bibinfo {volume} {127}},\
  \bibinfo {pages} {176401} (\bibinfo {year} {2021})}\BibitemShut {NoStop}%
\bibitem{YangLiuFang21} Jian Yang, Zheng-Xin Liu, Chen Fang, \href
  {https://arxiv.org/abs/2105.12738}{arXiv:2105.12738.}


  
\bibitem{sm} For details see the Supplemental Material at [url], which
includes Ref.~\cite{Bouhon-2021PRB, PAW-1994, PAW-1999, Cms-1996, tnc-1996, GGA-1996, wannier-1997, wannier-2001,Wu-2018}.


\bibitem [{\citenamefont {Chen}\ \emph {et~al.}(2021)\citenamefont {Chen},
  \citenamefont {Wang}, \citenamefont {Li},\ and\ \citenamefont
  {Sanyal}}]{Chen-2021}%
  \BibitemOpen
  \bibfield  {author} {\bibinfo {author} {\bibfnamefont {X.}~\bibnamefont
  {Chen}}, \bibinfo {author} {\bibfnamefont {D.}~\bibnamefont {Wang}}, \bibinfo
  {author} {\bibfnamefont {L.}~\bibnamefont {Li}},\ and\ \bibinfo {author}
  {\bibfnamefont {B.~S.}\ \bibnamefont {Sanyal}},\ }\href
  {https://arxiv.org/abs/2104.07390} {\bibfield  {journal} {\bibinfo
  {journal} {arXiv:2104.07390}}}\BibitemShut {NoStop}
\bibitem [{\citenamefont {Izmaylov}\ \emph {et~al.}(2006)\citenamefont
  {Izmaylov}, \citenamefont {Brothers},\ and\ \citenamefont
  {Scuseria}}]{HSE-2006}%
  \BibitemOpen
  \bibfield  {author} {\bibinfo {author} {\bibfnamefont {A.~F.}\ \bibnamefont
  {Izmaylov}}, \bibinfo {author} {\bibfnamefont {E.~N.}\ \bibnamefont
  {Brothers}},\ and\ \bibinfo {author} {\bibfnamefont {G.~E.}\ \bibnamefont
  {Scuseria}},\ }\bibfield  {title} {\bibinfo {title} {Linear-scaling
  calculation of static and dynamic polarizabilities in hartree-fock and
  density functional theory for periodic systems},\ }\href
  {https://doi.org/10.1063/1.2404667} {\bibfield  {journal} {\bibinfo
  {journal} {The Journal of Chemical Physics}\ }\textbf {\bibinfo {volume}
  {125}},\ \bibinfo {pages} {224105} (\bibinfo {year} {2006})},\ \Eprint
  {https://arxiv.org/abs/https://doi.org/10.1063/1.2404667}
  {https://doi.org/10.1063/1.2404667} \BibitemShut {NoStop}%
\bibitem [{\citenamefont {Bouhon}\ \emph {et~al.}(2021)\citenamefont {Bouhon},
  \citenamefont {Lange},\ and\ \citenamefont {Slager}}]{Bouhon-2021PRB}%
  \BibitemOpen
  \bibfield  {author} {\bibinfo {author} {\bibfnamefont {A.}~\bibnamefont
  {Bouhon}}, \bibinfo {author} {\bibfnamefont {G.~F.}\ \bibnamefont {Lange}},\
  and\ \bibinfo {author} {\bibfnamefont {R.-J.}\ \bibnamefont {Slager}},\
  }\bibfield  {title} {\bibinfo {title} {Topological correspondence between
  magnetic space group representations and subdimensions},\ }\href
  {https://doi.org/10.1103/PhysRevB.103.245127} {\bibfield  {journal} {\bibinfo
   {journal} {Phys. Rev. B}\ }\textbf {\bibinfo {volume} {103}},\ \bibinfo
  {pages} {245127} (\bibinfo {year} {2021})}\BibitemShut {NoStop}%
\bibitem [{\citenamefont {Bl\"ochl}(1994)}]{PAW-1994}%
  \BibitemOpen
  \bibfield  {author} {\bibinfo {author} {\bibfnamefont {P.~E.}\ \bibnamefont
  {Bl\"ochl}},\ }\bibfield  {title} {\bibinfo {title} {Projector augmented-wave
  method},\ }\href {https://doi.org/10.1103/PhysRevB.50.17953} {\bibfield
  {journal} {\bibinfo  {journal} {Phys. Rev. B}\ }\textbf {\bibinfo {volume}
  {50}},\ \bibinfo {pages} {17953} (\bibinfo {year} {1994})}\BibitemShut
  {NoStop}%
\bibitem [{\citenamefont {Kresse}\ and\ \citenamefont
  {Joubert}(1999)}]{PAW-1999}%
  \BibitemOpen
  \bibfield  {author} {\bibinfo {author} {\bibfnamefont {G.}~\bibnamefont
  {Kresse}}\ and\ \bibinfo {author} {\bibfnamefont {D.}~\bibnamefont
  {Joubert}},\ }\bibfield  {title} {\bibinfo {title} {From ultrasoft
  pseudopotentials to the projector augmented-wave method},\ }\href
  {https://doi.org/10.1103/PhysRevB.59.1758} {\bibfield  {journal} {\bibinfo
  {journal} {Phys. Rev. B}\ }\textbf {\bibinfo {volume} {59}},\ \bibinfo
  {pages} {1758} (\bibinfo {year} {1999})}\BibitemShut {NoStop}%
\bibitem [{\citenamefont {Kresse}\ and\ \citenamefont
  {Furthmüller}(1996)}]{Cms-1996}%
  \BibitemOpen
  \bibfield  {author} {\bibinfo {author} {\bibfnamefont {G.}~\bibnamefont
  {Kresse}}\ and\ \bibinfo {author} {\bibfnamefont {J.}~\bibnamefont
  {Furthmüller}},\ }\bibfield  {title} {\bibinfo {title} {Efficiency of
  ab-initio total energy calculations for metals and semiconductors using a
  plane-wave basis set},\ }\href
  {https://doi.org/https://doi.org/10.1016/0927-0256(96)00008-0} {\bibfield
  {journal} {\bibinfo  {journal} {Computational Materials Science}\ }\textbf
  {\bibinfo {volume} {6}},\ \bibinfo {pages} {15} (\bibinfo {year}
  {1996})}\BibitemShut {NoStop}%
\bibitem [{\citenamefont {Kresse}\ and\ \citenamefont
  {Furthm\"uller}(1996)}]{tnc-1996}%
  \BibitemOpen
  \bibfield  {author} {\bibinfo {author} {\bibfnamefont {G.}~\bibnamefont
  {Kresse}}\ and\ \bibinfo {author} {\bibfnamefont {J.}~\bibnamefont
  {Furthm\"uller}},\ }\bibfield  {title} {\bibinfo {title} {Efficient iterative
  schemes for ab initio total-energy calculations using a plane-wave basis
  set},\ }\href {https://doi.org/10.1103/PhysRevB.54.11169} {\bibfield
  {journal} {\bibinfo  {journal} {Phys. Rev. B}\ }\textbf {\bibinfo {volume}
  {54}},\ \bibinfo {pages} {11169} (\bibinfo {year} {1996})}\BibitemShut
  {NoStop}%
\bibitem [{\citenamefont {Perdew}\ \emph {et~al.}(1996)\citenamefont {Perdew},
  \citenamefont {Burke},\ and\ \citenamefont {Ernzerhof}}]{GGA-1996}%
  \BibitemOpen
  \bibfield  {author} {\bibinfo {author} {\bibfnamefont {J.~P.}\ \bibnamefont
  {Perdew}}, \bibinfo {author} {\bibfnamefont {K.}~\bibnamefont {Burke}},\ and\
  \bibinfo {author} {\bibfnamefont {M.}~\bibnamefont {Ernzerhof}},\ }\bibfield
  {title} {\bibinfo {title} {Generalized gradient approximation made simple},\
  }\href {https://doi.org/10.1103/PhysRevLett.77.3865} {\bibfield  {journal}
  {\bibinfo  {journal} {Phys. Rev. Lett.}\ }\textbf {\bibinfo {volume} {77}},\
  \bibinfo {pages} {3865} (\bibinfo {year} {1996})}\BibitemShut {NoStop}%
\bibitem [{\citenamefont {Marzari}\ and\ \citenamefont
  {Vanderbilt}(1997)}]{wannier-1997}%
  \BibitemOpen
  \bibfield  {author} {\bibinfo {author} {\bibfnamefont {N.}~\bibnamefont
  {Marzari}}\ and\ \bibinfo {author} {\bibfnamefont {D.}~\bibnamefont
  {Vanderbilt}},\ }\bibfield  {title} {\bibinfo {title} {Maximally localized
  generalized wannier functions for composite energy bands},\ }\href
  {https://doi.org/10.1103/PhysRevB.56.12847} {\bibfield  {journal} {\bibinfo
  {journal} {Phys. Rev. B}\ }\textbf {\bibinfo {volume} {56}},\ \bibinfo
  {pages} {12847} (\bibinfo {year} {1997})}\BibitemShut {NoStop}%
\bibitem [{\citenamefont {Souza}\ \emph {et~al.}(2001)\citenamefont {Souza},
  \citenamefont {Marzari},\ and\ \citenamefont {Vanderbilt}}]{wannier-2001}%
  \BibitemOpen
  \bibfield  {author} {\bibinfo {author} {\bibfnamefont {I.}~\bibnamefont
  {Souza}}, \bibinfo {author} {\bibfnamefont {N.}~\bibnamefont {Marzari}},\
  and\ \bibinfo {author} {\bibfnamefont {D.}~\bibnamefont {Vanderbilt}},\
  }\bibfield  {title} {\bibinfo {title} {Maximally localized wannier functions
  for entangled energy bands},\ }\href
  {https://doi.org/10.1103/PhysRevB.65.035109} {\bibfield  {journal} {\bibinfo
  {journal} {Phys. Rev. B}\ }\textbf {\bibinfo {volume} {65}},\ \bibinfo
  {pages} {035109} (\bibinfo {year} {2001})}\BibitemShut {NoStop}%
\bibitem [{\citenamefont {Wu}\ \emph {et~al.}(2018)\citenamefont {Wu},
  \citenamefont {Zhang}, \citenamefont {Song}, \citenamefont {Troyer},\ and\
  \citenamefont {Soluyanov}}]{Wu-2018}%
  \BibitemOpen
  \bibfield  {author} {\bibinfo {author} {\bibfnamefont {Q.}~\bibnamefont
  {Wu}}, \bibinfo {author} {\bibfnamefont {S.}~\bibnamefont {Zhang}}, \bibinfo
  {author} {\bibfnamefont {H.-F.}\ \bibnamefont {Song}}, \bibinfo {author}
  {\bibfnamefont {M.}~\bibnamefont {Troyer}},\ and\ \bibinfo {author}
  {\bibfnamefont {A.~A.}\ \bibnamefont {Soluyanov}},\ }\bibfield  {title}
  {\bibinfo {title} {Wanniertools: An open-source software package for novel
  topological materials},\ }\href
  {https://doi.org/https://doi.org/10.1016/j.cpc.2017.09.033} {\bibfield
  {journal} {\bibinfo  {journal} {Computer Physics Communications}\ }\textbf
  {\bibinfo {volume} {224}},\ \bibinfo {pages} {405} (\bibinfo {year}
  {2018})}\BibitemShut {NoStop}%
\end{thebibliography}%

\end{document}


\preprint{APS/123-QED}

\title{Supplemental Materials: Quantum Anomalous Hall Effect in Antiferromagnetism}

\author{Peng-Jie Guo}
\affiliation{Department of Physics and Beijing Key Laboratory of Opto-electronic Functional Materials $\&$ Micro-nano Devices, Renmin University of China, Beijing 100872, China}
\author{Zheng-Xin Liu}
\affiliation{Department of Physics and Beijing Key Laboratory of Opto-electronic Functional Materials $\&$ Micro-nano Devices, Renmin University of China, Beijing 100872, China}
\author{Zhong-Yi Lu}
\affiliation{Department of Physics and Beijing Key Laboratory of Opto-electronic Functional Materials $\&$ Micro-nano Devices, Renmin University of China, Beijing 100872, China}
\noaffiliation

\date{\today}


\maketitle

\section{Manipulation of  The Weyl Cones}

In this section, we firstly turn off SOC and investigate the number of Weyl cones when tuning the value of the staggered chemical potential $\mu_-$. Then we analyze the symmetry constraints under which the cones can be gapped out by tuning on SOC. The constraints of the symmetry groups will be helpful in the search of AFM materials to realize the QAH effect. 

\subsection{Symmetry protected Band crossing and the Number of Weyl cones}

Here we analyze the conditions for the existence of Weyl cones of the model (1) in the main text. As mentioned in the main text, the term  $\Gamma^{12}_k=0$ on the boundary of the BZ where $k_x=\pi$ or $k_y=\pi$.  Therefore, on the BZ boundary the hybridization between two species of fermions $C_{1k}$ and $C_{2k}$ vanishes owing to the protection of their different quantum numbers under certain symmetries (see Sec.\ref{sec:appIB} for details). Consequently,  if the band inversion between the $C_{1k}$ and the $C_{2k}$ bands takes place on the BZ boundary, then the appearance of Weyl cones will be protected by the corresponding symmetry. 


\begin{figure*}[htbp]
\centering
\!\!\! \!\!\! \!\!\! \!\!\! \subfigure[$m=0.1$]{ \includegraphics[width=6cm]{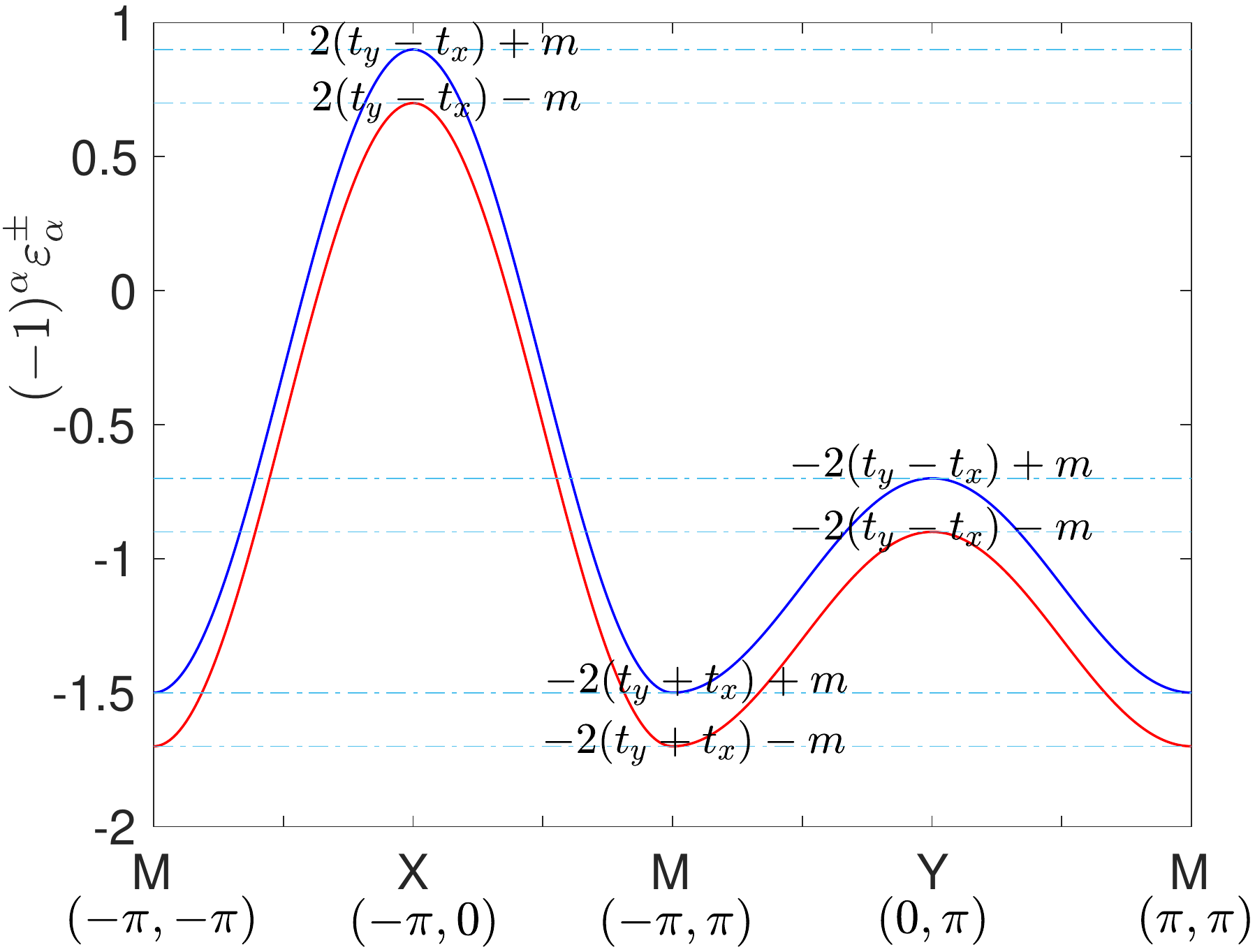}\label{fig:SM1a}}
\!\!\! \!\!\! \!\!\! \subfigure[$m=0.4$]{ \includegraphics[width=6cm]{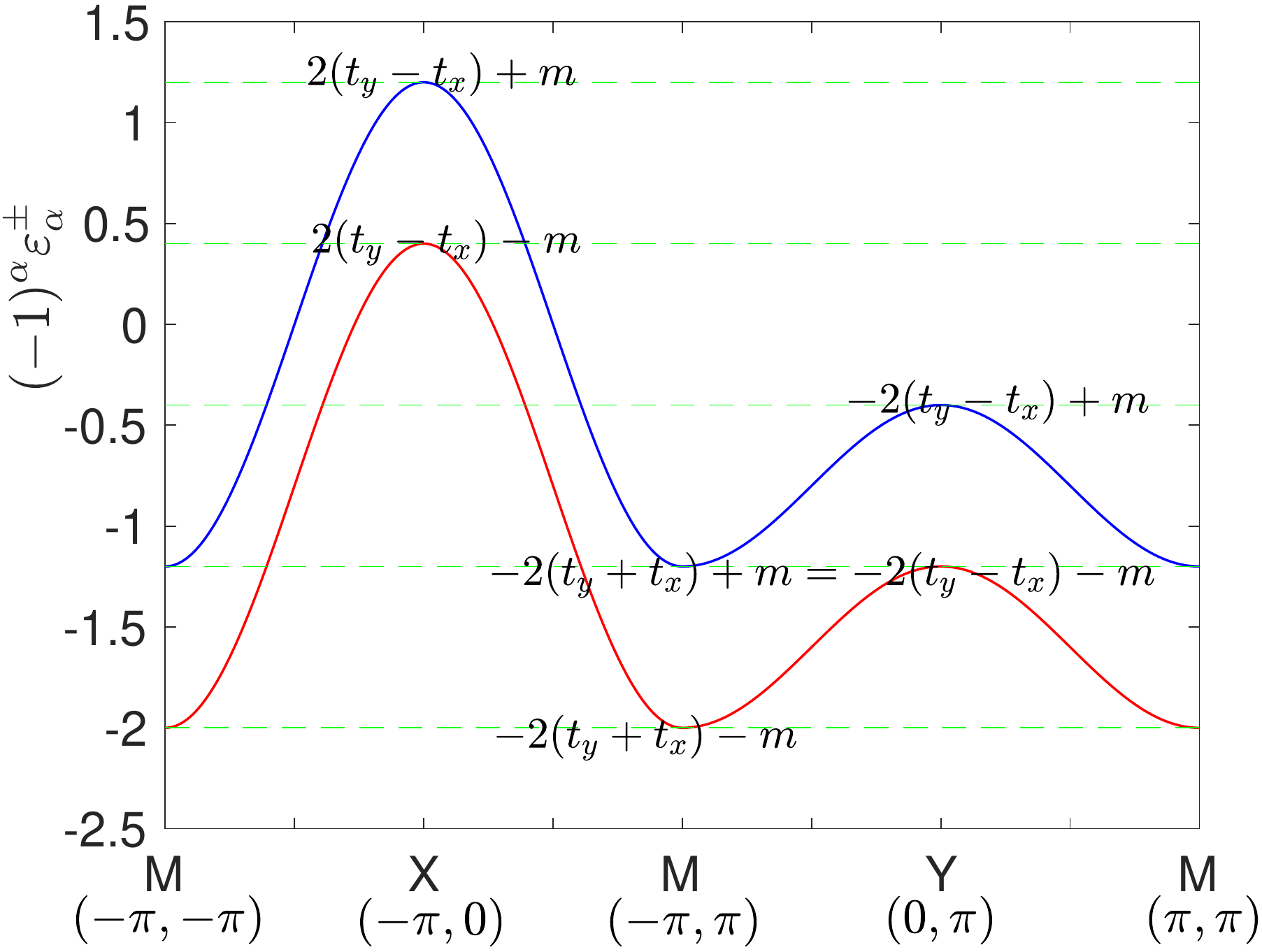}\label{fig:SM1b}}
\!\!\! \!\!\! \!\!\! \subfigure[$m=0.8$]{\includegraphics[width=6cm]{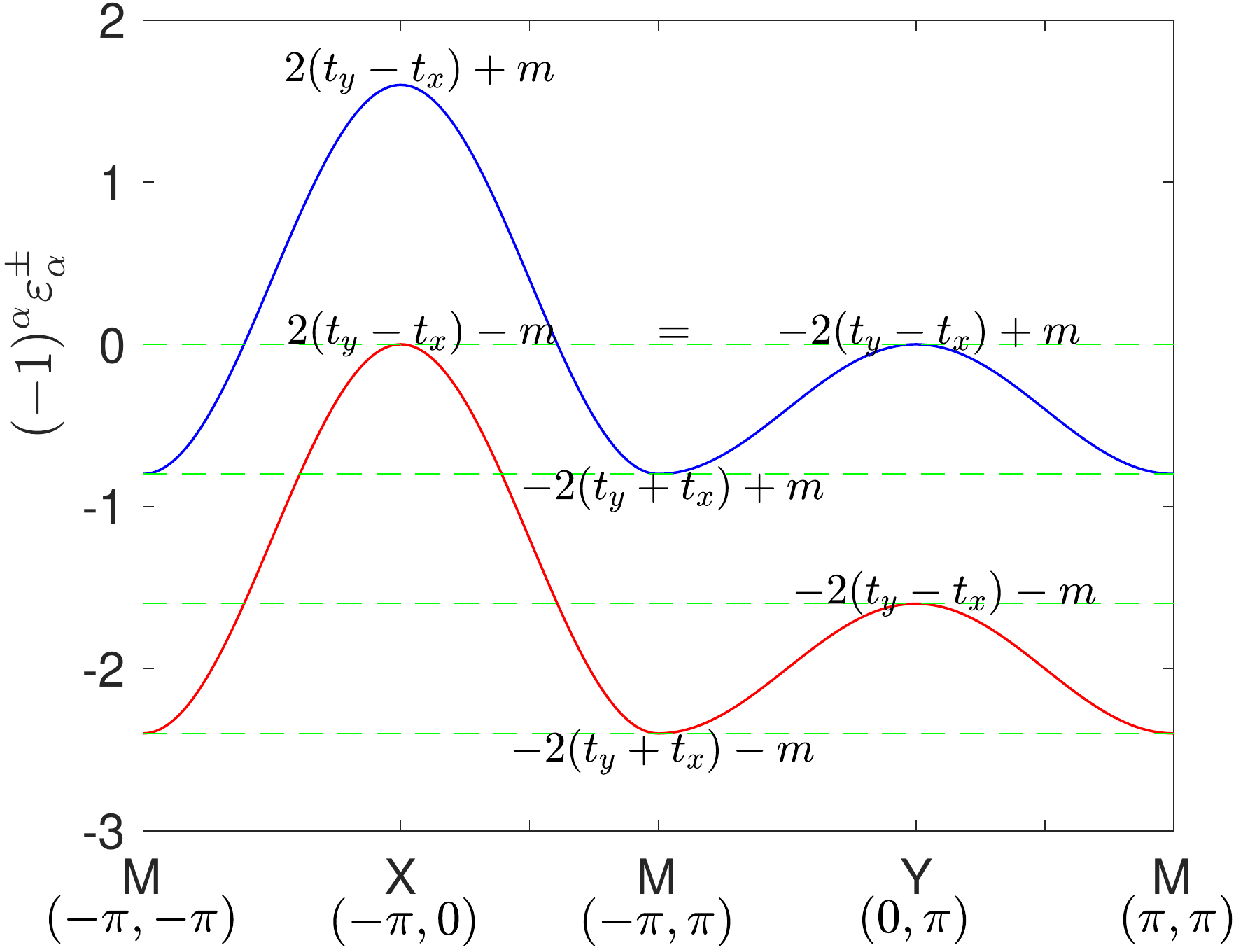}\label{fig:SM1c}}
\!\!\! \!\!\! \!\!\! \subfigure[$m=1.2$]{\includegraphics[width=6cm]{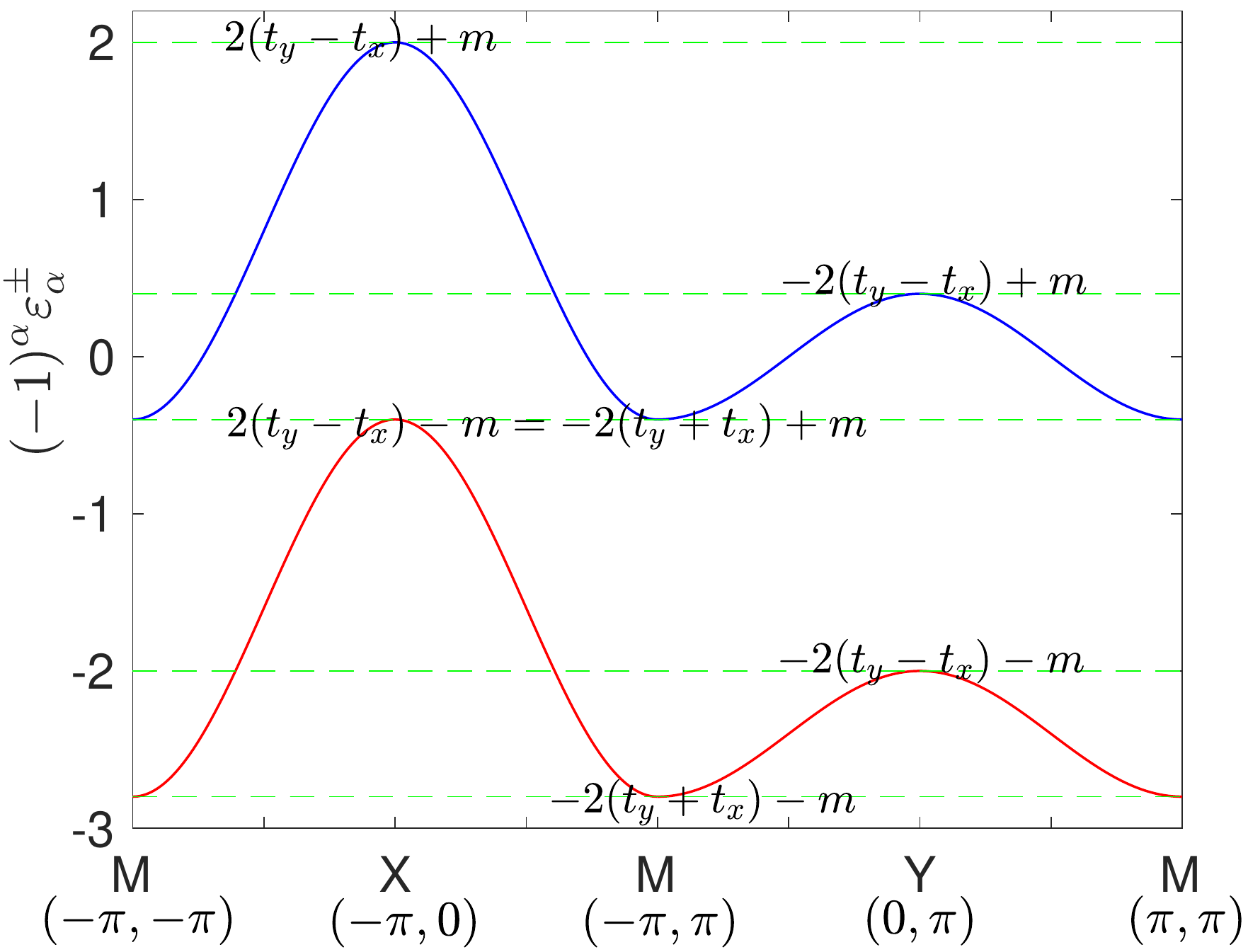}\label{fig:SM1d}}
\!\!\! \!\!\! \!\!\! \subfigure[$m=1.6$]{\includegraphics[width=6cm]{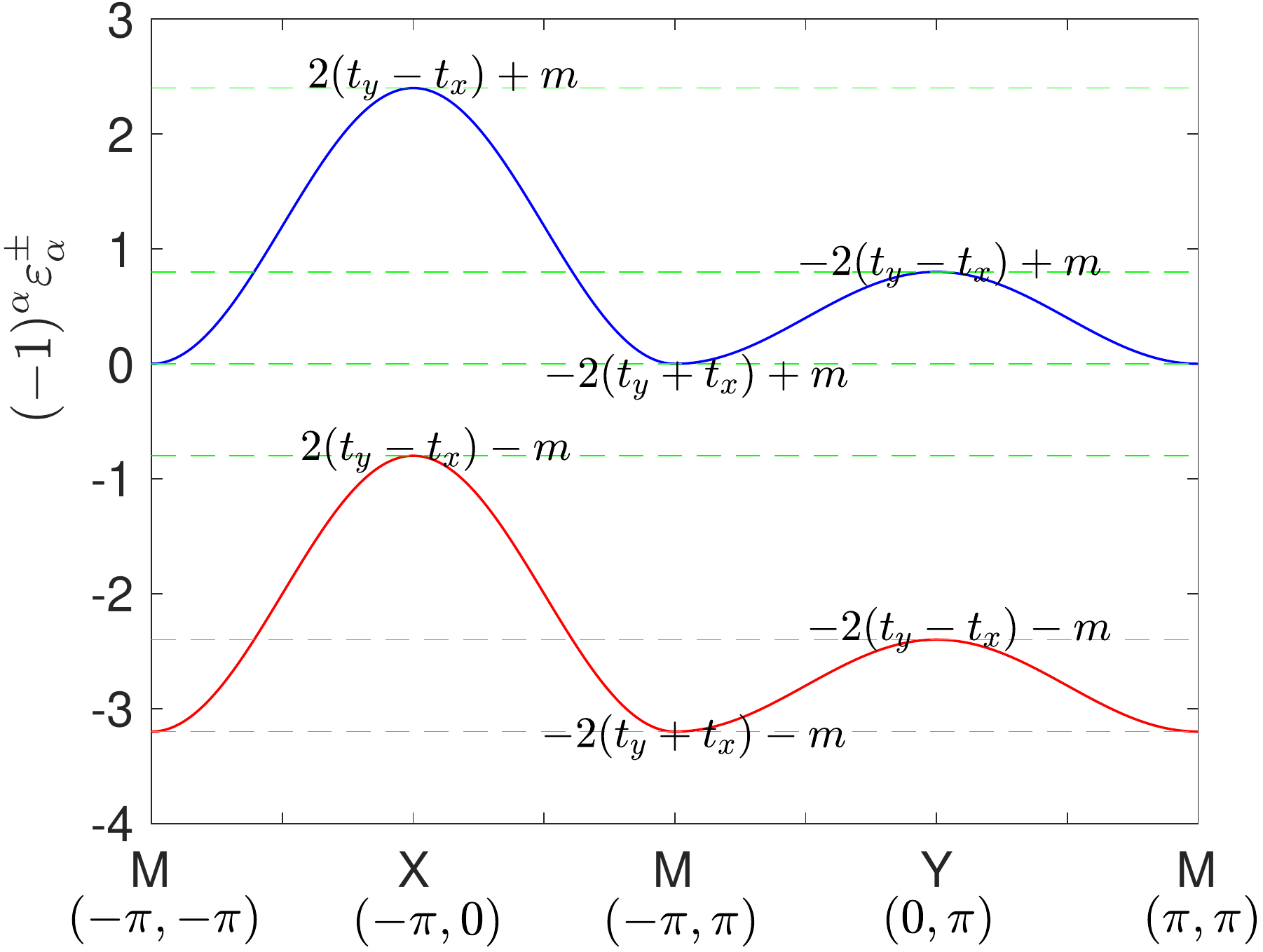}\label{fig:SM1e}}
	\caption{ The bands of $(-1)^\alpha \varepsilon^\pm_\alpha(k)$ with fixed $t_y=0.6, t_x=0.2$ are plotted in (a)$\sim$(e) with the increasing of $m=|\pmb m|$. The number of times that these bands crossed by the level $\mu_-$ determines the number of Weyl cones.}\label{fig:SM1}
\end{figure*}

On the BZ boundary the energy spectrum is determined by $\Gamma^{(\alpha)}_k$, whose eigenvalues read 
$$E_{\alpha}^\pm(k) = (-1)^\alpha(2t\cos k_x+2t_y\cos k_y \pm m  -\mu_-).$$ 
The band crossing can be tuned by the staggered chemical potential $\mu_- ={1\over2}(\mu_1-\mu_2)$.   To see this, we define the following quantity, 
\Beq
\varepsilon_\alpha^\pm(k) = (-1)^\alpha (2t_x\cos {k_x} +2t_y\cos k_y \pm m),
\Eeq
and consequently $$(-1)^\alpha \varepsilon_\alpha^\pm(k) =  2t_x\cos {k_x} +2t_y\cos k_y \pm m,$$ which is independent on $\alpha$.

The number of Weyl cones can be read out from the times that the bands of $(-1)^\alpha\varepsilon_\alpha^\pm$ (see Fig.\ref{fig:SM1} for illustration) are crossed by the level of $\mu_-$. In the following we analyze the dependence of the cone number and $\mu_-$ in details.

Without lose of generality, we fix $t_x, t_y$ and assume $t_y-t_x>0$ (the case with $t_y-t_x<0$ is very similar). Then we vary the value of $m$, and look into the range of $\mu_-$ within which the number of Weyl cones is stable. It turns out that there are four different situations, as enumerated below.

\begin{itemize}
\item [Case (1)] $2(t_y-t_x)-m>-2(t_y-t_x)+m>-2(t_y-t_x)-m>-2(t_y+t_x)+m$, see Fig.\ref{fig:SM1a}\\ 
A),  if $\mu_- \in \big(+2(t_y-t_x)-m, +2(t_y-t_x)+m\big)$, $\mu_-$ cross the bands twice, there are two (1 pair of) Weyl cones;\\
B),  if $\mu_- \in \big(-2(t_y-t_x)+m, +2(t_y-t_x)-m\big)$, $\mu_-$ cross the bands four times, there are four (2 pairs of) Weyl cones;\\
C),  if $\mu_- \in \big(-2(t_y-t_x)-m, -2(t_y-t_x)+m\big)$,  there are six (3 pairs of) Weyl cones;\\
D),  if $\mu_- \in \big(-2(t_y+t_x)+m, -2(t_y-t_x)-m\big)$,  there are eight (4 pairs of) Weyl cones;\\
E),  if $\mu_- \in \big(-2(t_y+t_x)-m, -2(t_y+t_x)-m\big)$,  there are four (2 pairs of) Weyl cones;\\
F),  otherwise, there are zero Weyl cone. 

\item [Case (2)] $2(t_y-t_x)-m>-2(t_y-t_x)+m>-2(t_y+t_x)+m>-2(t_y-t_x)-m$,  see Fig.\ref{fig:SM1b} \& \ref{fig:SM1c}\\ 
A),  if $\mu_- \in \big(+2(t_y-t_x)-m, +2(t_y-t_x)+m\big)$, there are two (1 pair of) Weyl cones;\\
B),  if $\mu_- \in \big(-2(t_y-t_x)+m, +2(t_y-t_x)-m\big)$, there are four (2 pairs of) Weyl cones;\\
C),  if $\mu_- \in \big(-2(t_y+t_x)+m, -2(t_y-t_x)+m\big)$,  there are six (3 pairs of) Weyl cones;\\
D),  if $\mu_- \in \big(-2(t_y-t_x)-m, -2(t_y+t_x)+m\big)$,  there are two (1 pair of) Weyl cones;\\
E),  if $\mu_- \in \big(-2(t_y+t_x)-m, -2(t_y-t_x)-m\big)$,  there are four (2 pairs of) Weyl cones;\\
F),  otherwise, there are zero Weyl cone. 

\item [Case (3)] $-2(t_y-t_x)+m>2(t_y-t_x)-m>-2(t_y+t_x)+m>-2(t_y-t_x)-m$,  see Fig.\ref{fig:SM1c} \& \ref{fig:SM1d}\\
A),  if $\mu_- \in \big(-2(t_y-t_x)+m, +2(t_y-t_x)+m\big)$, there are two (1 pair of) Weyl cones;\\
B),  if $\mu_- \in \big(+2(t_y-t_x)-m, -2(t_y-t_x)+m\big)$, there are four (2 pairs of) Weyl cones;\\
C),  if $\mu_- \in \big(-2(t_y+t_x)+m, +2(t_y-t_x)-m\big)$,  there are six (3 pairs of) Weyl cones;\\
D),  if $\mu_- \in \big(-2(t_y-t_x)-m, -2(t_y+t_x)+m\big)$,  there are two (1 pair of) Weyl cones;\\
E),  if $\mu_- \in \big(-2(t_y+t_x)-m, -2(t_y-t_x)-m\big)$,  there are four (2 pairs of) Weyl cones;\\
F),  otherwise, there are zero Weyl cone. 

\item [Case (4)] $-2(t_y-t_x)+m>-2(t_y+t_x)+m>2(t_y-t_x)-m>-2(t_y-t_x)-m$,  see Fig.\ref{fig:SM1e}\\
A),  if $\mu_- \in \big(-2(t_y-t_x)+m, +2(t_y-t_x)+m\big)$, there are two (1 pair of) Weyl cones;\\
B),  if $\mu_- \in \big(-2(t_y+t_x)+m,-2(t_y-t_x)+m\big)$, there are four (2 pairs of) Weyl cones;\\
C),  if $\mu_- \in \big(-2(t_y-t_x)-m, +2(t_y-t_x)-m\big)$,  there are two (1 pair of) Weyl cones;\\
D),  if $\mu_- \in \big(-2(t_y+t_x)-m, -2(t_y-t_x)-m\big)$,  there are four (2 pairs of) Weyl cones;\\
E),  otherwise, there are zero Weyl cone.
\end{itemize}
When $\mu_-$ equals the `transition' points, some pairs of cones merge into quadratic touches at the X point $(\pi,0)$, or the Y point $(0,\pi)$, or the M point $(\pi,\pi)$. \\

Among all the above cases, when $$\big|m-2|t_y-t_x|\big|< |\mu_-| <m+2|t_y-t_x|$$ is satisfied, namely, when 
$\big|m-2|t_y-t_x|\big|<  \mu_-  <m+2|t_y-t_x|$ or $-m-2|t_y-t_x|< \mu_- <-\big|m-2|t_y-t_x|\big|$, there are an odd number of pairs of cones. Actually, this is also a necessary condition for the nonzero Chern number in our model.

\subsection{Symmetry VS. the mass of the Weyl cones}\label{sec:appIB}

In this section we study the symmetry conditions under which the Weyl cones can be gapped out. The constraints of the symmetry groups will be helpful to search for AFM materials in realizing the QAH effect. 

\subsubsection{\bf Without SOC.} Firstly, we turn off SOC and analyze the quantum numbers that protect the Weyl cones.  We assume that the two bases $\hat x$ and $\hat y$  of the lattice vectors are orthogonal such that the crystalline point group is $D_{2h}$ generated by $C_2^x, C_2^y$ and $\mathcal I$.  Without SOC, the spin rotation symmetry operation and lattice rotation symmetry operation are unlocked, and the resultant point symmetry group is a spin point group (see the main text) which is generated by $(E||C_2^x), (E||C_2^y), (E||\mathcal I)$ and $(C_2^\perp\mathcal T||\mathcal T), (C_2^{\pmb m} || E)$. Here we have adopted the notation $(g||h)$ to denote the spin point group element as a combined operation of spin rotation $g$ and lattice operation $h$.

To illustrate, we analyze the boundary line $(k_x, \pi)$ which is invariant under the $(E||C_2^x)$ symmetry. Under the action of $(E||C_2^x)$, the fermion $C_{1}$ on site 1 is transformed into itself, but the fermion $C_2$ on site 2 is shifted by a lattice constant $\pmb y = a_2\hat y$ (here we assume $a_1,a_2$ are the lattice constants). Therefore, the fermion $C_{1k}$ carries quantum number 1 of $(E||C_2^x)$, while $C_{2k}$ carries a quantum number $e^{-i\pi}=-1$ remembering that $k_y=\pi$. These different quantum numbers protect the band crossing points, i.e. the Weyl cones,  from being gapped out. Since $(E,\mathcal I)$ is always a symmetry, the symmetry that providing the quantum numbers can also be chosen as $(E||M_x) = (E||C_2^x)\times (E,\mathcal I)$. Similarly, the Weyl cones on the $(\pi, k_y)$ boundary line, if exist, are protected by the quantum numbers of the $(E||C_2^y)$ or $(E||M_y)$ symmetry. 

However,  the Weyl cones may be robust against being gapped out even with the absence of the symmetry elements (such as $(E||C_2^x)$) which provide the quantum numbers. For instance, if the crystalline symmetry is lowered by applying shear strain, the Weyl cones will still be stable. 

Actually, the robustness of the cones are owing to the combined anti-unitary symmetry operation $(C_2^\perp \mathcal T||\mathcal {IT}) = (E||\mathcal I)\times(C_2^\perp \mathcal T||\mathcal T)$. This symmetry $(C_2^\perp \mathcal T||\mathcal {IT})$ is always there since the inversion symmetry $(E||\mathcal I)$ is preserved as we require and that the AFM order we are considering is collinear such that $(C_2^\perp \mathcal T||\mathcal T)$ is a symmetry. This spin point group symmetry operation $(C_2^\perp \mathcal T||\mathcal {IT})$ differs with the familiar space-time inversion operation $(\mathcal T||\mathcal {IT})$ (usually denoted as $\mathcal {IT}$ for simplicity) by a pure spin rotation $(C_2^\perp||E)$.  Notice that the space-time inversion symmetry $(\mathcal T||\mathcal {IT})$ is a symmetry of momentum points in the whole BZ, and it ensures that the Berry curvature is vanishing for all momentum point $k$ whose energy spectrum is gapped. Furthermore, in the presence of $(\mathcal T||\mathcal {IT})$ symmetry, the Berry phase for a closed loop is quantized to $\pi$ or $0$: it is equal to $\pi$ if it encloses an odd number of Weyl cones and equals to zero if it encloses an even number of Weyl cones. On the other hand, the spin rotation $(C_2^\perp||E)$ does not act on the momentum $k$ and does not change the Berry curvature. Consequently, the above results for the symmetry $(\mathcal T||\mathcal {IT})$ are also valid for the spin-point symmetry $(C_2^\perp \mathcal T||\mathcal {IT})$.

Since $(C_2^\perp \mathcal T||\mathcal {IT})$ ensures that the Berry phase of any loop in the BZ is quantized to $\pi$ or $0$ (here we assume that the spectrum on the loop is fully gapped), the Weyl cones cannot be solely gapped out. Resultantly, under perturbations (such as shear strain) preserving the $(C_2^\perp \mathcal T||\mathcal {IT})$ symmetry, the Weyl cones adiabatically shift their positions. The number of cones can be reduced in pair if two cones move toward each other, then merge and finally open a gap. So if the $(C_2^\perp \mathcal T||\mathcal {IT})$ symmetry preserving perturbations are not  strong, then the Weyl cones remain robust. 

\subsubsection{\bf With SOC.} 
When SOC is present, the physical symmetry of a magnetically ordered system is described by a magnetic group where the lattice rotation operations are locked with the corresponding spin rotations. 

{\bf In-plane AFM order.} Firstly we consider in-plane AFM order. Without loss of generality, we assume that $\pmb m\parallel\hat y$. In this case, $(C_2^y || C_2^y)$ is a symmetry operation if the crystalline point group contains $C_2^y$. Since $(C_2^y || C_2^y)$ has quantum numbers $\pm i$,  these quantum numbers 
can still protect the band crossing on the line $(\pi,k_y)$ (whose little co-group contains the symmetry operation $(C_2^y || C_2^y)$) from being gapped out.  In our model, the Hamiltonian on the line $(\pi,k_y)$ reads 
\Beq
H_k = C_{1 k}^\dag\Big[ -4i \pmb \lambda \cdot \pmb\sigma \sin {k_y\over2} \Big] C_{2 k} + \sum_{\alpha} (-1)^\alpha C_{\alpha k}^\dag  \Big[ 2t_y \cos k_y - 2t_x + \pmb m\cdot \pmb\sigma - \mu_-\Big] C_{\alpha k}. 
\Eeq
Since the $(C_2^y || C_2^y)$ symmetry is a physical requirement, only $\lambda_x$ and $\lambda_z$ components are symmetry allowed:  the lattice part $(E||C_2^y)$ transforms $\pmb \lambda^{d_1}$ to  $\pmb \lambda^{d_2}=- \pmb \lambda^{d_1}$ and the spin part $ (C_2^y||E)$ send $- \lambda^{d_1}_{x,z}$ back to $\lambda^{d_1}_{x,z}$. In this case, the Weyl cons on the line $(\pi,k_y)$, if exist, are still robust even if the $C_{1k}$ and $C_{2k}$ fermions are indeed hybridized owing to the SOC.  

\begin{figure*}[htbp]
\centering
\includegraphics[width=15cm]{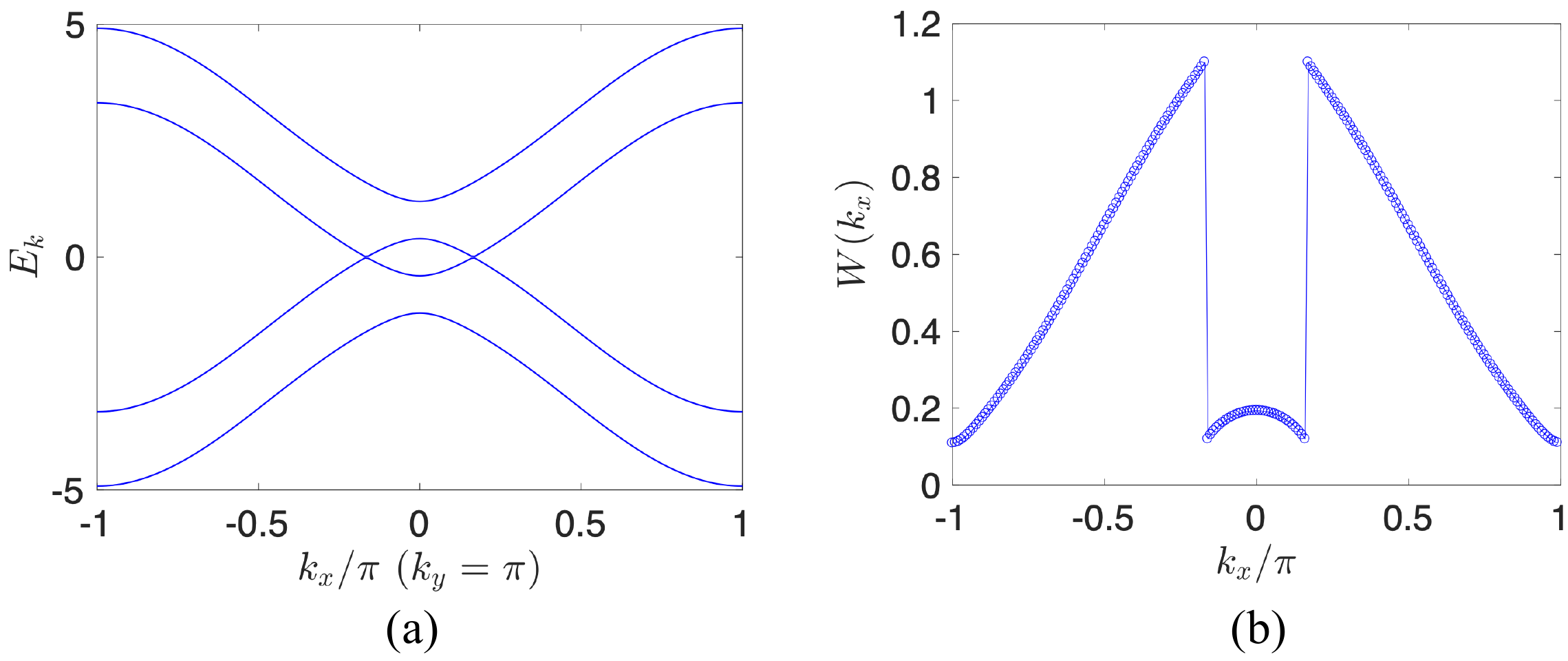}
	\caption{ (a) The Weyl cones are robust against SOC when $\lambda_y\neq0, \lambda_z\neq0$ and $\lambda_x=0$ given that $\pmb m\parallel \hat x$. (b) The 1D chain with given $k_x$ is not topological since the Berry phase $W(k_x)$ (see Eq.~(\ref{1D})) is not quantized.}\label{fig:SM2}
\end{figure*}

Similarly, the quantum numbers $\pm i$ of $(C_2^y||M_y) = (C_2^y||C_2^y)\times (E||\mathcal I)$ can protect the Weyl cones on the line $(k_x,\pi)$ from being gapped out (see Fig.~\ref{fig:SM2}(a)). Notice that the symmetry $(C_2^y||M_y)$ is preserved as long as the SOC term $\pmb \lambda$ lies in the $xz$-plane such that $\pmb m\cdot\pmb\lambda=0$. 

Therefore, if $\pmb m\cdot\pmb\lambda=0$ then the Weyl cones on the BZ boundary are still robust owing to the protection of quantum numbers of either $(C_2^y || C_2^y)$  or $(C_2^y||M_y)$. This result indicates that if the AFM order is parallel to an in-plane $C_2$ axis, then the Weyl cones resulting from band inversion will remain robust and QAH effect cannot be realized.





If $\pmb\lambda\cdot\pmb m\neq0$, then all the Weyl cones obtain a mass, which results in a QAH insulator. If both $\pmb\lambda$ and $\pmb m$ are parallel to $\hat y$, then the symmetry group of the model reads $2'/m'=\{E,\mathcal I, C_2^x\mathcal T, M_x\mathcal T\}$. In this case, the crystalline point group is $C_{2h}$. If $\pmb \lambda$ is neither perpendicular nor parallel to $\pmb m$, then the symmetry group of the model reduces to $\bar 1= \{E,\mathcal I\}$. Therefore, QAH insulators can be realized in in-plane AFM ordered systems if the lattice structure is either triclinic or monoclinic. 

{\bf Out-of-plane AMF order.} Then we consider the case where the AFM order is perpendicular to the lattice plane, namely $\pmb m\parallel \hat z$. In this case, the QAH effect can be realized if $\lambda_z\neq 0$.

If $\pmb\lambda \parallel\hat z$, then the model has a higher symmetry which is described by the magnetic point group $m'm'm=\{E, C_2^x\mathcal T, C_2^y\mathcal T, C_2^z, \mathcal I, M_x\mathcal T,  M_y\mathcal T, M_z\}$. Therefore, QAH insulator can be realized in AFM materials with crystalline point group $D_{2h}$ generated by $C_2^x, C_2^y$ and $\mathcal I$.

The above symmetry conditions can be released. For instance, if $\pmb\lambda$ is deviated from $\hat z$ or if the lattice axis $\hat a$ and $\hat b$ are not orthogonal, then the magnetic point group can be lowered to $2'/m'=\{E, \mathcal I, C_2^x\mathcal T,   M_x\mathcal T\}$ or $\bar 1=\{E,\mathcal I\}$. Therefore, if the magnetic order is perpendicular to the lattice plane,  QAH insulators can be possibly found in AFM materials having triclinic, monoclinic, or orthogonal lattice structure. 

If the unit cell contains more than two magnetic sites, the resultant Chern number maybe greater than one\cite{Bouhon-2021PRB}.  

%

It should be mentioned that our discussion is restricted to the mechanism that the Weyl cones come from the inversion of bands with different quantum numbers. The Weyl cones may also arise from symmetry protection or from critical points\cite{Jiang-2018}, in which cases the above symmetry analysis is no-longer applicable. 

\section{Chern number and Edge states}

When the two  Weyl cones on the boundary $(k_x, \pi)$ are gapped out, one obtains a QAH insulator because the Chern number is nonzero. Here we briefly summarize the method in the computation of the Chern number in discretized momentum space. 

Firstly we calculate the Berry connection $A_{i}(k)\in [-\pi, \pi)$ with $i=x,y$ and $k=(k_x, k_y)$, 
\beq\label{BerryA}
A_i(k) = {\rm Im} \big(\log \det \mathcal O_i(k)\big),
\eeq  
where ${\rm Im}$ means the imaginary part and $O_i(k)$ is a 2 by 2 matrix defined from the following overlap
\Beq
[\mathcal O_i(k)]^{ab} = \langle \phi_a(k) | \phi_b(k+\delta k_i)\rangle,
\Eeq
where  $a,b=1,2$ are the indices of the occupied bands, $\delta k_i$ is the step length of the discretized momenta along the $i$-direction, and $|\phi_a(k)\rangle$ are eigenvectors of $H_k$ with $$H_k|\phi_a(k)\rangle =E_k|\phi_a(k)\rangle.$$    

As $\Gamma^{12}_k$ contains the variables ${k_{x}\over 2}, {k_{y}\over 2}$, the Hamiltonian is not periodic in the first BZ and the same for their eigenstates. Nevertheless, the Chern number can still be computed, as long as both $|\phi(k_x,-\pi)\rangle$ and $|\phi(k_x,\pi)\rangle$, and similarly both $|\phi(-\pi,k_y)\rangle$ and $|\phi(\pi,k_y)\rangle$ are kept.
With these states, the following matrices are defined,
\Beq
&&[\mathcal O_x(\pi-\delta k_x, k_y)]^{ab} = \langle \phi_a(\pi-\delta_x, k_y) | \phi_b(\pi, k_y)\rangle,\ \ \
[\mathcal O_y(\pi,k_y)]^{ab} = \langle \phi_a(\pi, k_y) | \phi_b(\pi, k_y+\delta k_y)\rangle,\\
&&[\mathcal O_y(k_x,\pi-\delta k_y)]^{ab} = \langle \phi_a(k_x,\pi-\delta k_y) | \phi_b(k_x, \pi)\rangle,\ \ \ 
[\mathcal O_x(k_x,\pi)]^{ab} = \langle \phi_a(k_x,\pi) | \phi_b(k_x+\delta k_x,\pi)\rangle.
\Eeq
Hence $A_x(\pi-\delta k_x, k_y), A_y(\pi,k_y)$ and $A_y(k_x,\pi-\delta k_y), A_x(k_x, \pi)$ are obtained from (\ref{BerryA}). 

\begin{figure*}[htbp]
\centering
\includegraphics[width=15cm]{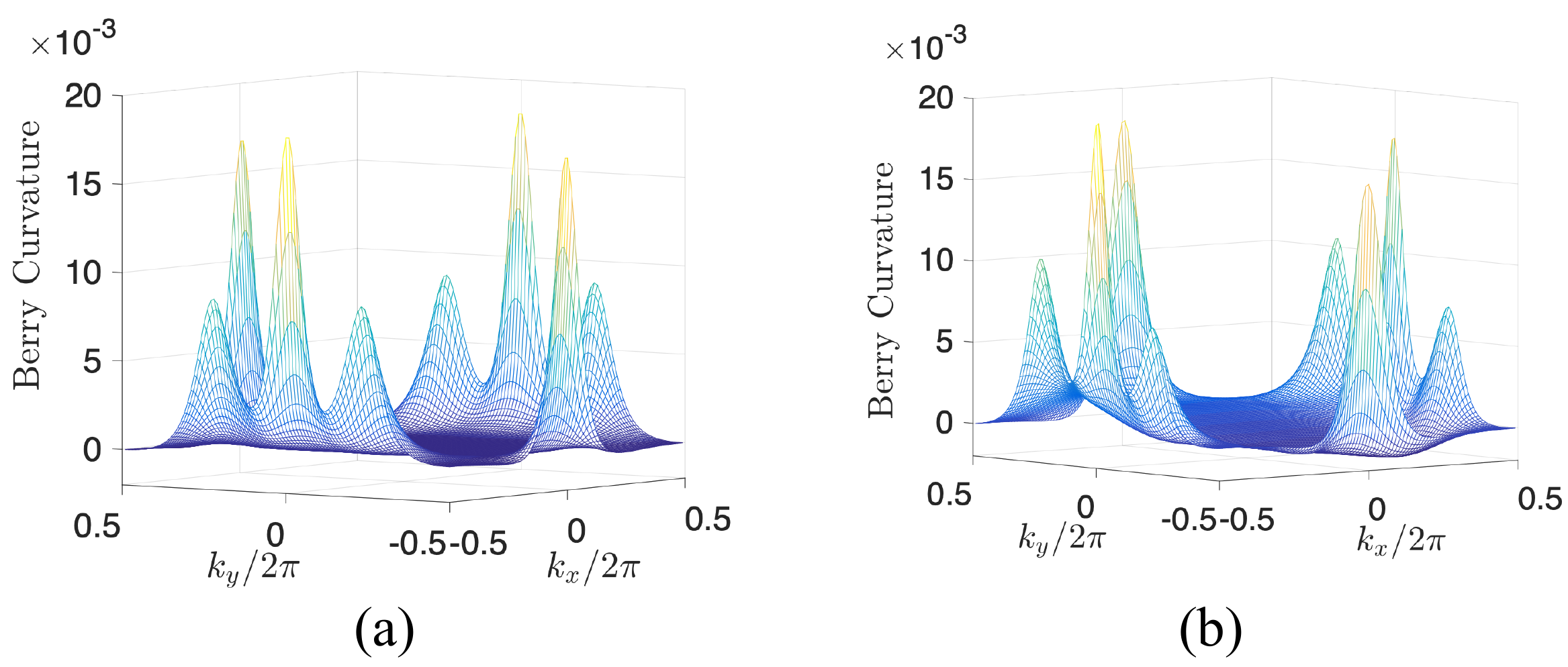}
	\caption{ Pictures of Berry curvature calculated in two different conventions both resulting in a Chern number $C=1$. (a) With the original $\pmb d_i$, the Berry curvature preserves the inversion symmetry; (b) With the transformation $\pmb d_i \to \pmb d_i - {1\over2}(\pmb x+\pmb y)$, the Berry curvature no longer preserves the inversion symmetry.}\label{fig:SMBerryCurv}
\end{figure*}

Then we calculate the Berry curvature $F(k)$ for $k_x\in[-\pi,\pi), k_y\in[-\pi,\pi)$,
\Beq
F(k) = \big[ A_x(k) + A_y(k+\delta k_x) - A_x(k+\delta k_y) - A_y(k)\big] {\rm mod\ } 2\pi.
\Eeq
When $k_x, k_y$ are fine-grained, $F_{ij}(k)$ should be a small number. However, owing to a random gauge choice of the numerically obtained $|\phi_j(k)\rangle$, sometimes $F_{ij}(k)$ may be as big as close to $2\pi N, N\in\mathbb Z$, so we adopt ${\rm mod\ } 2\pi$ to guarantee that $F_{ij}(k)$ is indeed small (See Fig.~\ref{fig:SMBerryCurv}).

Finally, we obtain the Chern number from the summation of $F$ over the BZ,
\beq\label{Chern}
C = {1\over 2\pi}   \sum_{k_x\in [-\pi,\pi)} \sum_{k_y\in[-\pi,\pi)} F(k_x,k_y) .
\eeq

\begin{figure*}[htbp]
\centering
\!\!\! \!\!\! \!\!\! \!\!\! \subfigure[Edge states of QAH insulator]{ \includegraphics[width=8cm]{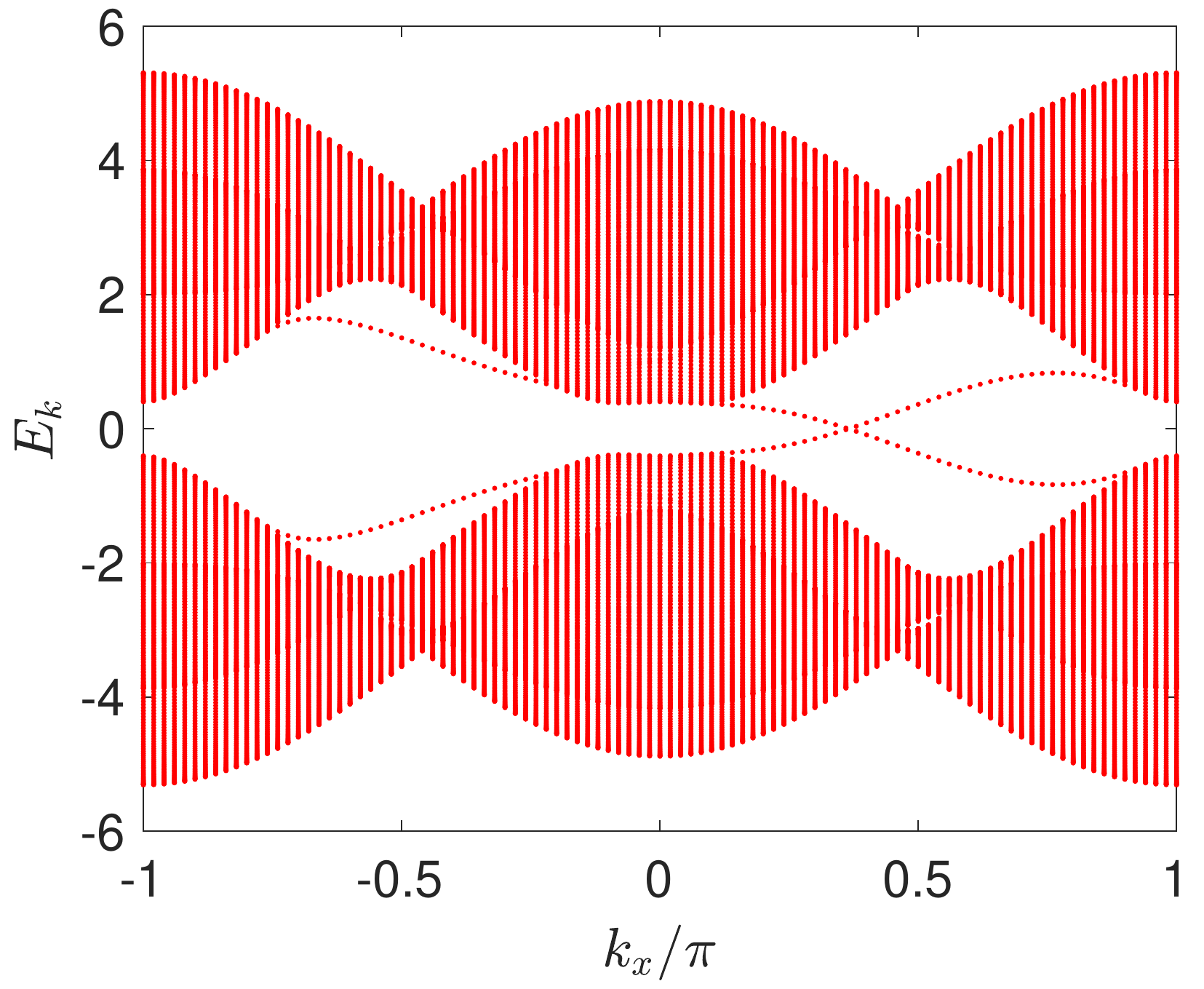}\label{fig:SM3a}}
\!\!\! \!\!\! \!\!\! \subfigure[Edge states of Weyl semimetal]{ \includegraphics[width=8cm]{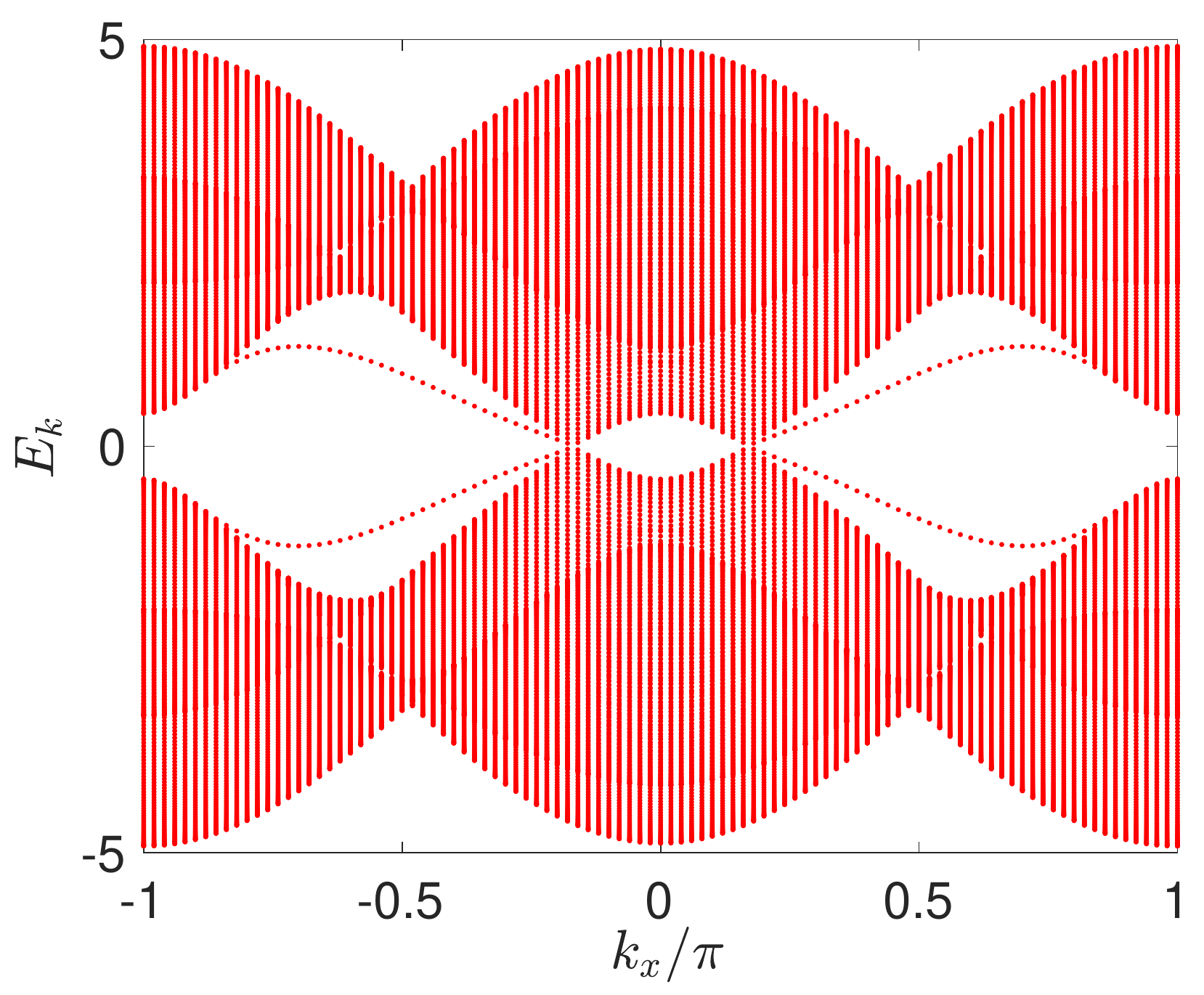}\label{fig:SM3b}}
	\caption{ Edge states of (a) QAH insulator with $\lambda_x=0.5$, (b) Weyl semimetal with $\lambda_x=0$. }\label{fig:SM3}
\end{figure*}

Above we illustrate that the Chern number can be calculated regardless that the Hamiltonian is not periodic in the first BZ. The reason is that the Hamiltonian can be made periodic in the first BZ under a nonsingular unitary transformation which does not change the Chern number. This can be done by adopting another convention to coordinate the lattice sites within a unit cell. When shifting $\pmb d_i \to \pmb d_i - {1\over2}(\pmb x+\pmb y)$, the new Hamiltonian no longer contains  ${k_{x}\over 2}, {k_{y}\over 2}$, thus are periodic in the first BZ. In this case, one can simply replace $|\phi_a(\pi,k_y)\rangle$ by $|\phi_a(-\pi,k_y)\rangle$ and replace $|\phi_a(k_x,\pi)\rangle$ by $|\phi_a(k_x,-\pi)\rangle$. Accordingly the formulas for   $A_x(\pi-\delta k_x, k_y), A_y(\pi,k_y)=A_y(-\pi,k_y)$ and $A_y(k_x,\pi-\delta k_y), A_x(k_x, \pi)=A_x(k_x, -\pi)$ will be modified, but the formula (\ref{Chern}) for  computing the Chern number remains unchanged. 

In momentum space, the transformation $\pmb d_i \to \pmb d_i - {1\over2}(\pmb x+\pmb y)$ is $k$-dependent, consequently the Berry curvature is affected (the Chern number remains the same). For instance, the Berry curvature is inversion symmetric in the first convention(see Fig.\ref{fig:SMBerryCurv} (a)), but after the transformation  $\pmb d_i \to \pmb d_i - {1\over2}(\pmb x+\pmb y)$, the resultant Berry curvature is no longer inversion symmetric (see Fig.\ref{fig:SMBerryCurv} (b)). 

The nontrivial bulk Chern number is further verified by the existence of chiral edge states. If one adopt the cylinder geometry with periodic boundary condition along the $x$-direction (such that $k_x$ is still a good quantum number) and open boundary condition along the $y$-direction, there will be two branches of chiral edge states, as shown in Fig.\ref{fig:SM3}(a).

In the following we discuss the edge states of the Weyl semimetal with $\lambda_x=0$. Suppose that the parameters are chosen such that there is one pair of Weyl cones in the BZ boundary line $(k_x,\pi)$. We still adopt cylinder geometry as before.
Since $k_x$ is a good quantum number, one can consider the Hamiltonian with fixed $k_x$ as a 1D chain. If the 1D chain has zero energy edge states, then there will exist zero-energy edge states on the open boundary of the 2D cylinder.

Since the zero edge mode of 1D chain requires a topological invariant or $\pi$-quantization of the Berry phase.
 Here we calculate the Zak phase of the 1D chain with given $k_x$,
\beq\label{1D}
W (k_x)= \Big [{1\over \pi} \sum_{k_y\in[-\pi,\pi)} A_y(k_x, k_y)  \Big]  {\rm mod\ 2}, 
\eeq 
where $ A_y(k_x, k_y)$ is defined in Eq.~(\ref{BerryA}).

Since quantization condition requires that the Hamiltonian is periodic in the 1D momentum space $k_x\in [-\pi,\pi]$,  we adopt the single valued Hamiltonian  after the transformation $\pmb d_i \to \pmb d_i - {1\over2}(\pmb x+\pmb y)$. The Berry phase indeed jumps by $\pi$ when $k_x$ passes the position of the Weyl cones, however the quantity $W(k_x)$ in (\ref{1D}) is not an integer for most $k_x$ (see Fig.~\ref{fig:SM2}(b)).  Therefore, we conclude that the invariant is absent and consequently  no zero-energy edge modes are protected. However, the jump of the Berry phase at the position of the Weyl cones indicates that the situations are indeed different inside the two Weyl cones and outside the Weyl cones. As shown in Fig.~\ref{fig:SM3}(b), the dispersive middle gap edge states are found outside the cones but are absent inside the cones.

\section{Details about the first principle calculatitions} 

The electronic structures of a monolayer CrO were studied with the projector augmented wave method\cite{PAW-1994, PAW-1999} as implemented in the VASP package\cite{Cms-1996, tnc-1996}. The Perdew-Burke-Ernzerhof exchange-correlation functional at the generalized gradient approximation level was adopted to describe the interaction between the ionic cores and the valence electrons\cite{GGA-1996}. For the electronic correlation in the $d$ orbitals of Cr$^{2+}$, we adopted the Hubbard $U$ formalism. The kinetic energy cutoff of the plane-wave basis was set to 700 eV and a 20 $\times$ 20 $\times$ 1 was adopted for the $k$-point mesh. The Gaussian smearing method with a width of 0.01 eV was utilized for the Fermi surface broadening. Both cell parameters and internal atomic positions were fully relaxed until the forces on all atoms were smaller than 0.001 eV/\AA{}. The calculated lattice parameters were 1.4 percent less than the previous values\cite{Chen-2021}. A 20 \AA{} vacuum layer was used to avoid the residual interaction between adjacent layers. The Berry curvature of monolayer CrO was calculated by using the Wannier90 package\cite{wannier-1997, wannier-2001}. The topological invariants and the chiral edge states of monolayer CrO were studied by WannierTools package\cite{Wu-2018}.

\bibliography{ref}